%% file: transits.tex
\documentclass[preprint2]{exoplanet}
\usepackage{times,amsmath}

\newcommand{\refs}{\par\noindent\hangindent=1pc\hangafter=1}
\begin{document}

\def\ltsima{$\; \buildrel < \over \sim \;$}
\def\lsim{\lower.5ex\hbox{\ltsima}}
\def\gtsima{$\; \buildrel > \over \sim \;$}
\def\gsim{\lower.5ex\hbox{\gtsima}}

\title{\textbf{\LARGE Transits and Occultations}}

\author {\textbf{\large Joshua N.\ Winn}}
\affil{\small\em Massachusetts Institute of Technology}

\begin{abstract}
\begin{list}{ }
{\rightmargin 0.5in}
\baselineskip = 11pt
\parindent=1pc {\small When we are fortunate enough to view an
  exoplanetary system nearly edge-on, the star and planet periodically
  eclipse each other. Observations of eclipses---transits and
  occultations---provide a bonanza of information that cannot be
  obtained from radial-velocity data alone, such as the relative
  dimensions of the planet and its host star, as well as the
  orientation of the planet's orbit relative to the sky plane and
  relative to the stellar rotation axis. The wavelength-dependence of
  the eclipse signal gives clues about the the temperature and
  composition of the planetary atmosphere.  Anomalies in the timing or
  other properties of the eclipses may betray the presence of
  additional planets or moons. Searching for eclipses is also a
  productive means of discovering new planets. This chapter reviews
  the basic geometry and physics of eclipses, and summarizes the
  knowledge that has been gained through eclipse observations, as well
  as the information that might be gained in the future.
  \\~\\~\\~}
\end{list}
\end{abstract}  

\section{INTRODUCTION}

{\it From immemorial antiquity, men have dreamed of a royal road to
  success---leading directly and easily to some goal that could be
  reached otherwise only by long approaches and with weary toil.
  Times beyond number, this dream has proved to be a
  delusion.... Nevertheless, there are ways of approach to unknown
  territory which lead surprisingly far, and repay their followers
  richly. There is probably no better example of this than eclipses of
  heavenly bodies.} --- Henry Norris Russell (1948)

\vskip 0.1in

Vast expanses of scientific territory have been traversed by
exploiting the occasions when one astronomical body shadows
another. The timing of the eclipses of Jupiter's moons gave the first
accurate measure of the speed of light. Observing the passage of Venus
across the disk of the Sun provided a highly refined estimate of the
astronomical unit. Studying solar eclipses led to the discovery of
helium, the recognition that Earth's rotation is slowing down due to
tides, and the confirmation of Einstein's prediction for the
gravitational deflection of light. The analysis of eclipsing binary
stars---the subject Russell had in mind---enabled a precise
understanding of stellar structure and evolution.

\begin{figure*}
\centerline{\includegraphics[angle=-90, scale=0.55]{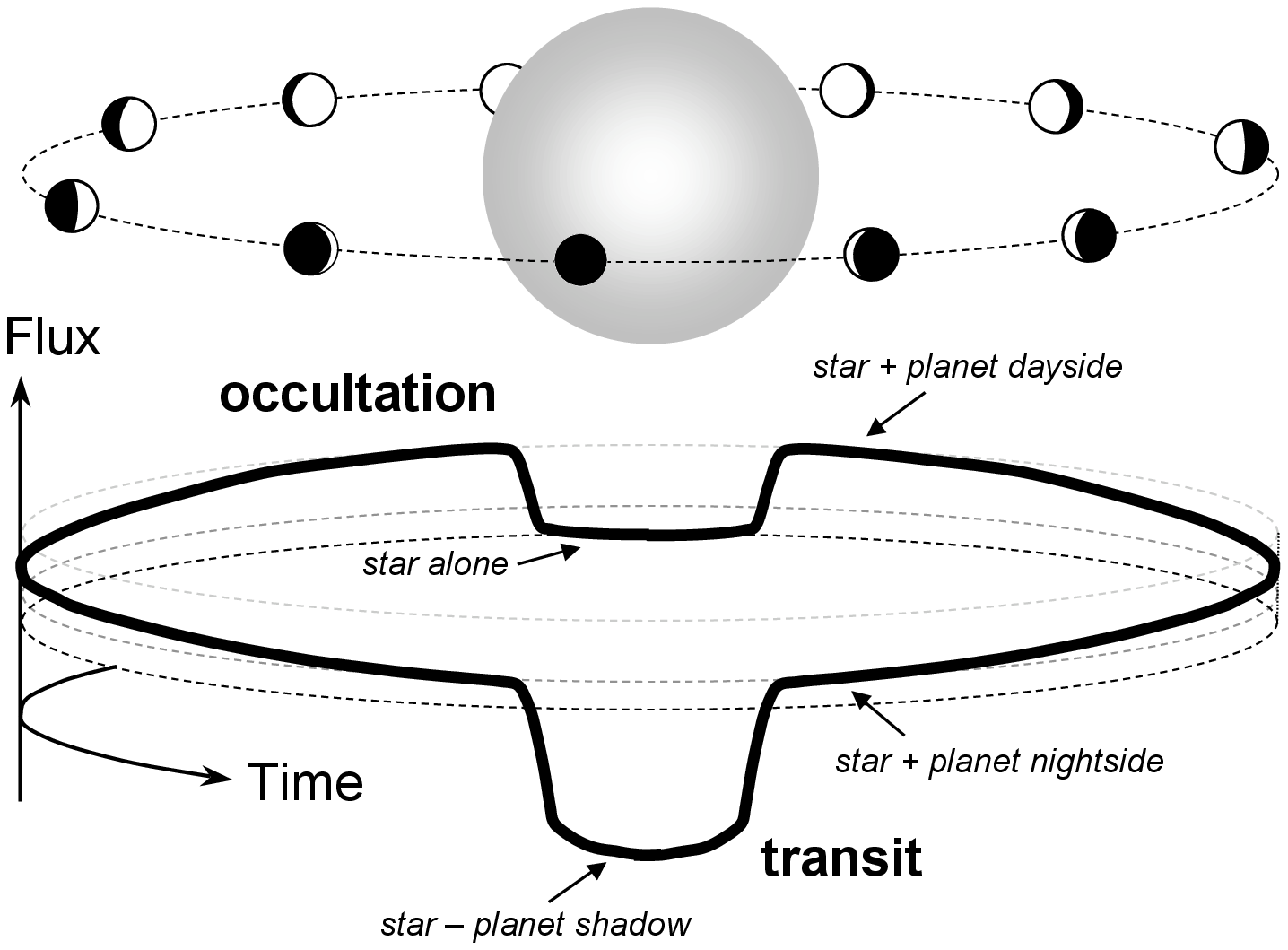}}
\caption{\small Illustration of transits and occultations. Only the
  combined flux of the star and planet is observed. During a transit,
  the flux drops because the planet blocks a fraction of the
  starlight. Then the flux rises as the planet's dayside comes into
  view. The flux drops again when the planet is occulted by the
  star. \label{fig:circular_diagram}}
\end{figure*}

Continuing in this tradition, eclipses are the ``royal road'' of
exoplanetary science. We can learn intimate details about exoplanets
and their parent stars through observations of their combined light,
without the weary toil of spatially resolving the planet and the star
(see Figure~\ref{fig:circular_diagram}). This chapter shows how
eclipse observations are used to gain knowledge of the planet's orbit,
mass, radius, temperature, and atmospheric constituents, along with
other details that are otherwise hidden. This knowledge, in turn,
gives clues about the processes of planet formation and evolution and
provides a larger context for understanding the properties of the
solar system.

An {\it eclipse} is the obscuration of one celestial body by
another. When the bodies have very unequal sizes, the passage of the
smaller body in front of the larger body is a {\it transit} and the
passage of the smaller body behind the larger body is an {\it
  occultation}. Formally, transits are cases when the full disk of the
smaller body passes completely within that of the larger body, and
occultations refer to the complete concealment of the smaller body.
We will allow those terms to include the {\it grazing} cases in which
the bodies' silhouettes do not overlap completely. Please be aware
that the exoplanet literature often refers to occultations as {\it
  secondary eclipses} (a more general term that does not connote an
extreme size ratio), or by the neologisms ``secondary transit'' and
``anti-transit.''

This chapter is organized as follows. Section~2 describes the geometry
of eclipses and provides the foundational equations, building on the
discussion of Keplerian orbits in the chapter by Murray and
Correia. Readers seeking a more elementary treatment involving only
circular orbits may prefer to start by reading
Sackett~(1999). Section~3 discusses many scientific applications of
eclipse data, including the determination of the mass and radius of
the planet. Section~4 is a primer on observing the apparent decline in
stellar brightness during eclipses (the {\it photometric}
signal). Section~5 reviews some recent scientific accomplishments, and
Section~6 offers some thoughts on future prospects.

\bigskip
\centerline{\textbf{ 2. ECLIPSE BASICS}}
\bigskip
\noindent
\textbf{ 2.1 Geometry of eclipses}
\bigskip

\begin{figure*}
  \epsscale{1.5}
  \centerline{\includegraphics[angle=0, scale=0.75]{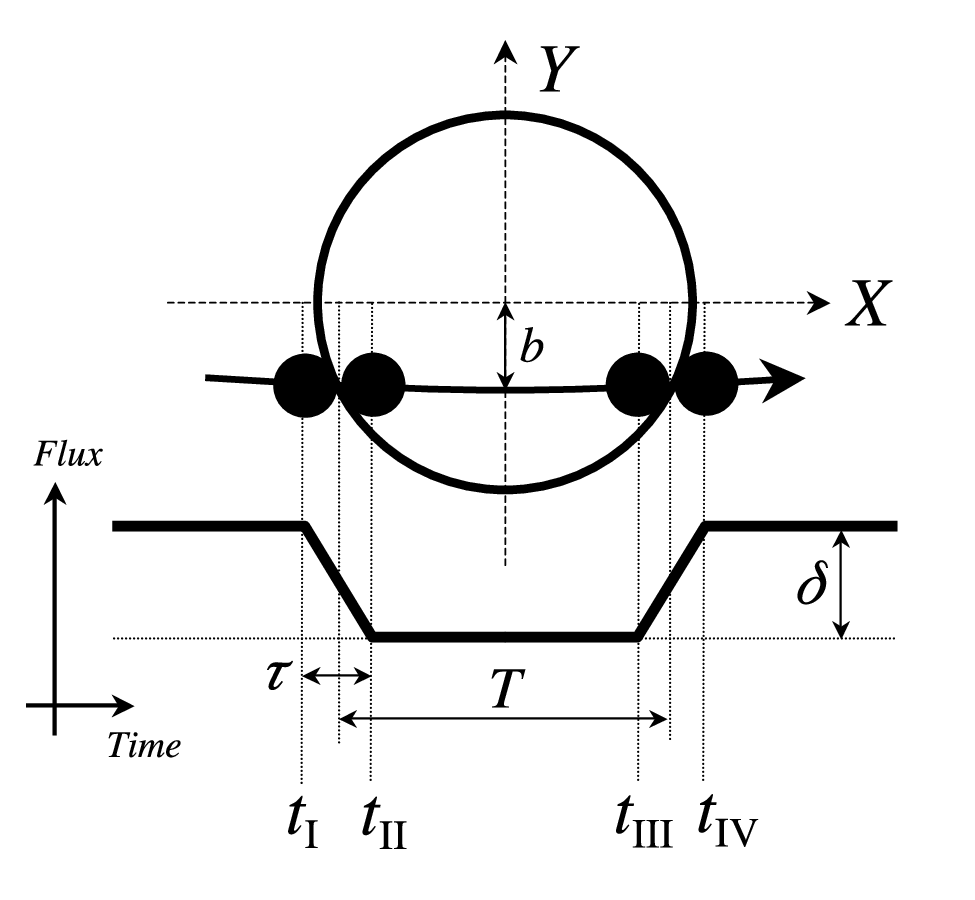}}
  \caption{\small Illustration of a transit, showing the coordinate
    system discussed in Section 2.1, the four contact points and the
    quantities $T$ and $\tau$ defined in Section 2.3, and the
    idealized light curve discussed in
    Section~2.4. \label{fig:transit_diagram}}
\end{figure*}

Consider a planet of radius $R_p$ and mass $M_p$ orbiting a star of
radius $R_\star$ and mass $M_\star$. The ratio $R_p/R_\star$ occurs
frequently enough to deserve its own symbol, for which we will use
$k$, in deference to the literature on eclipsing binary stars. As in
the chapter by Murray and Correia, we choose a coordinate system
centered on the star, with the sky in the $X$--$Y$ plane and the $+Z$
axis pointing at the observer (see
Figure~\ref{fig:transit_diagram}). Since the orientation of the line
of nodes relative to celestial north (or any other externally defined
axis) is usually unknown and of limited interest, we might as well
align the $X$ axis with the line of nodes; we place the descending
node of the planet's orbit along the $+X$ axis, giving
$\Omega=180^\circ$.

The distance between the star and planet is given by equation~(20) of
the chapter by Murray and Correia:
\begin{equation}
r = \frac{a(1-e^2)}{1+e\cos f},
\end{equation}
where $a$ is the semimajor axis of the relative orbit and $f$ is the
true anomaly, an implicit function of time depending on the orbital
eccentricity $e$ and period $P$ (see Section~3 of the chapter by
Murray and Correia). This can be resolved into Cartesian coordinates
using equations~(53-55) of the chapter by Murray and Correia, with
$\Omega=180^\circ$:
\begin{eqnarray}
X & = & -r \cos(\omega+f), \label{eq:x} \\
Y & = & -r \sin(\omega+f)\cos i, \label{eq:y} \\
Z & = & r \sin(\omega+f)\sin i. \label{eq:z}
\end{eqnarray}

If eclipses occur, they do so when $r_{\rm sky} \equiv \sqrt{X^2+Y^2}$
is a local minimum. Using equations~(\ref{eq:x}-\ref{eq:y}),
\begin{equation}
 r_{\rm sky} =
 \frac{a(1-e^2)}{1+e\cos f}
 \sqrt{ 1 - \sin^2(\omega + f) \sin^2 i }.
\label{eq:sky-projected-distance}
\end{equation}
Minimizing this expression leads to lengthy algebra (Kipping
2008). However, an excellent approximation that we will use throughout
this chapter is that eclipses are centered around {\it conjunctions},
which are defined by the condition $X=0$ and may be {\it inferior}
(planet in front) or {\it superior} (star in front). This gives
\begin{equation}
f_{{\rm tra}} = +\frac{\pi}{2} - \omega,~~~
f_{{\rm occ}} = -\frac{\pi}{2} - \omega,
\end{equation}
where here and elsewhere in this chapter, ``tra'' refers to transits
and ``occ'' to occultations. This approximation is valid for all cases
except extremely eccentric and close-in orbits with grazing eclipses.

The {\it impact parameter} $b$ is the sky-projected distance at
conjunction, in units of the stellar radius:
\begin{eqnarray}
b_{\rm tra} & = & \frac{a \cos i}{R_\star} \left(\frac{1-e^2}{1 + e\sin\omega}\right), \label{eq:b-tra} \\
b_{\rm occ} & = & \frac{a \cos i}{R_\star} \left(\frac{1-e^2}{1 - e\sin\omega}\right). \label{eq:b-occ}
\end{eqnarray}
For the common case $R_\star \ll a$, the planet's path across (or
behind) the stellar disk is approximately a straight line between the
points $X = \pm R_\star\sqrt{1-b^2}$ at $Y=bR_\star$.

\bigskip
\noindent
\textbf{ 2.2 Probability of eclipses}
\bigskip

\begin{figure*}
  \epsscale{1.0}
  \centerline{\includegraphics[angle=0, scale=0.6]{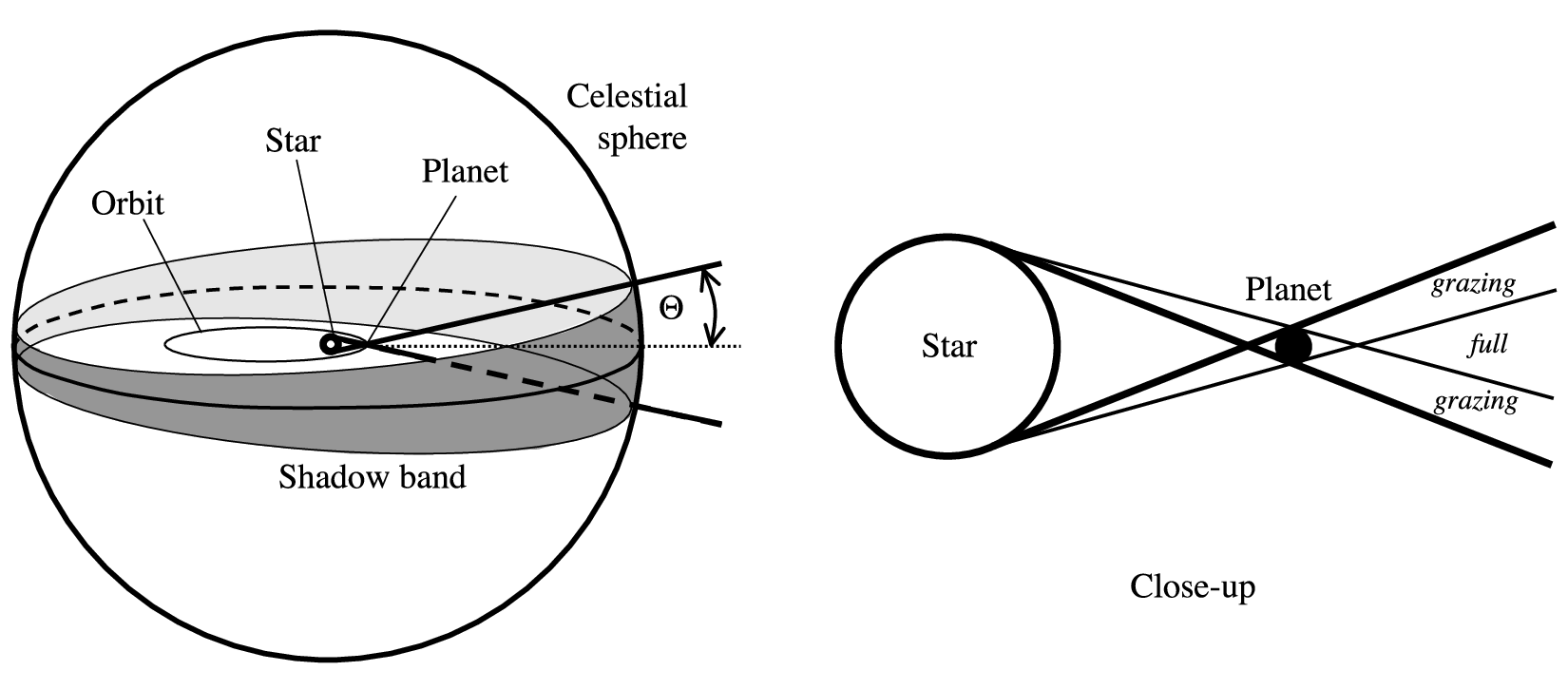}}
  \caption{\small Calculation of the transit probability. {\it
      Left.}---Transits are visible by observers within the penumbra
    of the planet, a cone with opening angle $\Theta$ with $\sin\Theta
    = (R_\star+R_p)/r$, where $r$ is the instantaneous star-planet
    distance. {\it Right.}---Close-up showing the penumbra (thick
    lines) as well as the antumbra (thin lines) within which the
    transits are full, as opposed to grazing.\label{fig:probcalc} }
\end{figure*}

Eclipses are seen only by privileged observers who view a planet's
orbit nearly edge-on. As the planet orbits its star, its shadow
describes a cone that sweeps out a band on the celestial sphere, as
illustrated in Figure~\ref{fig:probcalc}. A distant observer within
the shadow band will see transits. The opening angle of the cone,
$\Theta$, satisfies the condition $\sin\Theta = (R_\star+R_p)/r$ where
$r$ is the instantaneous star-planet distance. This cone is called the
{\it penumbra}. There is also an interior cone, the {\it antumbra},
defined by $\sin\Theta = (R_\star-R_p)/r$, inside of which the
transits are full (non-grazing).

A common situation is that $e$ and $\omega$ are known and $i$ is
unknown, as when a planet is discovered via the Doppler method (see
chapter by Lovis and Fischer) but no information is available about
eclipses. With reference to Figure~\ref{fig:probcalc}, the observer's
celestial longitude is specified by $\omega$, but the latitude is
unknown. The transit probability is calculated as the shadowed
fraction of the line of longitude, or more simply from the requirement
$|b| < 1 + k$, using equations~(\ref{eq:b-tra}-\ref{eq:b-occ}) and the
knowledge that $\cos i$ is uniformly distributed for a randomly-placed
observer. Similar logic applies to occultations, leading to the
results
\begin{eqnarray}
p_{\rm tra} & = & \left(\frac{R_\star \pm R_p}{a}\right)
                   \left(\frac{1 + e\sin\omega}{1 - e^2} \right),
\label{eq:transit-probability-ecc-argperi} \\
p_{\rm occ} & = & \left(\frac{R_\star \pm R_p}{a}\right)
                   \left(\frac{1 - e\sin\omega}{1 - e^2} \right),
\label{eq:occ-probability-ecc-argperi}
\end{eqnarray}
where the ``$+$'' sign allows grazing eclipses and the ``$-$'' sign
excludes them. It is worth committing to memory the results for the
limiting case $R_p \ll R_\star$ and $e=0$:
\begin{equation}
p_{\rm tra} = p_{\rm occ} = \frac{R_\star}{a}~\approx 0.005~\left( \frac{R_\star}{R_{\odot}} \right)
                        \left( \frac{a}{1~{\rm AU}} \right)^{-1}.
\end{equation}
For a circular orbit, transits and occultations always go together,
but for an eccentric orbit it is possible to see transits without
occultations or vice versa.

In other situations, one may want to marginalize over all possible
values of $\omega$, as when forecasting the expected number of
transiting planets to be found in a survey (see Section~4.1) or other
statistical calculations. Here, one can calculate the solid angle of
the entire shadow band and divide by $4\pi$, or average
equations~(\ref{eq:transit-probability-ecc-argperi}-\ref{eq:occ-probability-ecc-argperi})
over $\omega$, giving
\begin{equation}
p_{\rm tra} = p_{\rm occ} = \left(\frac{R_\star \pm R_p}{a}\right) \left(\frac{1}{1-e^2}\right).
\label{eq:transit-probability-ecc}
\end{equation}

Suppose you want to find a transiting planet at a particular orbital
distance around a star of a given radius. If a fraction $\eta$ of
stars have such planets, you must search at least $N \approx
(\eta~p_{\rm tra})^{-1}$ stars before expecting to find a transiting
planet. A sample of $>$200~$\eta^{-1}$ Sun-like stars is needed to
find a transiting planet at 1~AU. Close-in giant planets have an
orbital distance of approximately 0.05~AU and $\eta\approx 0.01$,
giving $N>$10$^3$ stars. In practice, many other factors affect the
survey requirements, such as measurement precision, time sampling, and
the need for spectroscopic follow-up observations (see Section 4.1).

\bigskip
\noindent
\textbf{ 2.3 Duration of eclipses}
\bigskip

In a non-grazing eclipse, the stellar and planetary disks are tangent
at four {\it contact times} $t_{\rm I}$--$t_{\rm IV}$, illustrated in
Figure~\ref{fig:transit_diagram}. (In a grazing eclipse, second and
third contact do not occur.) The {\it total} duration is $T_{\rm tot}
= t_{\rm IV}-t_{\rm I}$, the {\it full} duration is $T_{\rm full} =
t_{\rm III} - t_{\rm II}$, the {\it ingress} duration is $\tau_{\rm
  ing} = t_{\rm II}-t_{\rm I}$, and the {\it egress} duration is
$\tau_{\rm egr} = t_{\rm IV}-t_{\rm III}$.

Given a set of orbital parameters, the various eclipse durations can
be calculated by setting equation~(\ref{eq:sky-projected-distance})
equal to $R_\star \pm R_p$ to find the true anomaly at the times of
contact, and then integrating equation~(44) of the chapter by Murray
and Correia, e.g.,
\begin{equation}
t_{\rm III} - t_{\rm II} = \frac{P}{2\pi\sqrt{1-e^2}}
\int_{f_{\rm II}}^{f_{\rm III}} \left[\frac{r(f)}{a}\right]^2 \, df.
\end{equation}
For a circular orbit, some useful results are
\begin{equation}
T_{\rm tot} \equiv t_{\rm IV} - t_{\rm I} =
\frac{P}{\pi}\sin^{-1}\left[\frac{R_\star}{a}
  \frac{\sqrt{(1+k)^2-b^2}}{\sin i} \right],
\label{eq:t-tot}
\end{equation}
\begin{equation}
T_{\rm full} \equiv t_{\rm III} - t_{\rm II} =
\frac{P}{\pi}\sin^{-1}\left[\frac{R_\star}{a}
  \frac{\sqrt{(1-k)^2-b^2}}{\sin i} \right].
\label{eq:t-full}
\end{equation}
For eccentric orbits, good approximations are obtained by multiplying
equations~(\ref{eq:t-tot}-\ref{eq:t-full}) by
\begin{equation}
  \frac{\dot{X}(f_c)~[e=0]}{\dot{X}(f_c)} = \frac{\sqrt{1-e^2}}{1 \pm e\sin\omega}, \label{eq:xdot-ecc}
\end{equation}
a dimensionless factor to account for the altered speed of the planet
at conjunction. Here, ``$+$'' refers to transits and ``$-$'' to
occultations. One must also compute $b$ using the
eccentricity-dependent equations~(\ref{eq:b-tra}-\ref{eq:b-occ}).

For an eccentric orbit, $\tau_{\rm ing}$ and $\tau_{\rm egr}$ are
generally unequal because the projected speed of the planet varies
between ingress and egress. In practice the difference is slight; to
leading order in $R_\star/a$ and $e$,
\begin{equation}
\frac{\tau_e - \tau_i}{\tau_e + \tau_i} \sim
e\cos\omega \left( \frac{R_\star}{a} \right)^3 \left(1-b^2\right)^{3/2},
\end{equation}
which is $<$$10^{-2}~e$ for a close-in planet with $R_\star/a = 0.2$,
and even smaller for more distant planets. For this reason we will use
a single symbol $\tau$ to represent either the ingress or egress
duration. Another important timescale is $T \equiv T_{\rm tot} -
\tau$, the interval between the halfway points of ingress and egress
(sometimes referred to as contact times 1.5 and 3.5).

In the limits $e \rightarrow 0$, $R_p \ll R_\star \ll a$, and $b \ll
1-k$ (which excludes near-grazing events), the results are greatly
simplified:
\begin{equation}
T \approx T_0 \sqrt{1-b^2},~~
\tau \approx \frac{T_o k}{\sqrt{1-b^2}}, \label{eq:duration-simple}
\end{equation}
where $T_0$ is the characteristic time scale
\begin{eqnarray}
T_0 \equiv \frac{R_\star P}{\pi a}
 \approx 13~{\rm hr}~\left( \frac{P}{1~{\rm yr}} \right)^{1/3}
                \left( \frac{\rho_\star}{\rho_\odot} \right)^{-1/3}.
\label{eq:duration-mean-density}
\end{eqnarray}
For eccentric orbits, the additional factor given by
equation~(\ref{eq:xdot-ecc}) should be applied. Note that in deriving
equation~(\ref{eq:duration-mean-density}), we used Kepler's third law
and the approximation $M_p \ll M_\star$ to rewrite the expression in
terms of the stellar mean density $\rho_\star$. This is a hint that
eclipse observations give a direct measure of $\rho_\star$, a point
that is made more explicit in Section 3.1.

\bigskip
\noindent
\textbf{ 2.4 Loss of light during eclipses}
\bigskip

The combined flux $F(t)$ of a planet and star is plotted in
Figure~\ref{fig:circular_diagram}. During a transit, the flux drops
because the planet blocks a fraction of the starlight. Then the flux
rises as the planet's dayside comes into view. The flux drops again
when the planet is occulted by the star. Conceptually we may dissect
$F(t)$ as
\begin{equation}
F(t) = F_\star(t) + F_p(t) -
\begin{cases}
k^2 \alpha_{\rm tra}(t) F_\star(t)  & \text{transits,} \\
0 & \text{outside eclipses,} \\
\alpha_{\rm occ}(t) F_p(t)  & \text{occultations.}
\end{cases}
\end{equation}
where $F_\star,F_p$ are the fluxes from the stellar and planetary
disks, and the $\alpha$'s are dimensionless functions of order unity
depending on the overlap area between the stellar and planetary
disks. In general $F_\star$ may vary in time due to flares, rotation
of star spots and plages, rotation of the tidal bulge raised by the
planet, or other reasons, but for simplicity of discussion we take it
to be a constant. In that case, only the ratio $f(t) \equiv
F(t)/F_\star$ is of interest. If we let $I_p$ and $I_\star$ be the
disk-averaged intensities of the planet and star, respectively, then
$F_p/F_\star = k^2 I_p/I_\star$ and
\begin{equation}
f(t) = 1 + k^2 \frac{I_p(t)}{I_\star} -
\begin{cases}
k^2 \alpha_{\rm tra}(t) & \text{transits,} \\
0 & \text{outside eclipses,} \\
k^2 \frac{I_p(t)}{I_\star} \alpha_{\rm occ}(t) & \text{occultations.}
\end{cases}
\end{equation}

Time variations in $I_p$ are caused by the changing illuminated
fraction of the planetary disk (its {\it phase function}), as well as
any changes intrinsic to the planetary atmosphere. To the extent that
$I_p$ is constant over the relatively short timespan of a single
eclipse, all of the observed time variation is from the $\alpha$
functions. As a starting approximation the $\alpha$'s are trapezoids,
and $f(t)$ is specified by the depth $\delta$, duration $T$, ingress
or egress duration $\tau$, and time of conjunction $t_c$, as shown in
Figure~\ref{fig:transit_diagram}. For transits the maximum loss of
light is
\begin{equation}
  \delta_{\rm tra} \approx k^2~\left[1 - \frac{I_p(t_{\rm tra})}{I_\star}\right],
\end{equation}
and in the usual case when the light from the planetary nightside is
negligible, $\delta_{\rm tra} \approx k^2$. For occultations,
\begin{equation}
  \delta_{\rm occ} \approx k^2 \frac{I_p(t_{\rm occ})}{I_\star}.
\end{equation}

In the trapezoidal approximation the flux variation during ingress and
egress is linear in time. In reality this is not true, partly because
of the nonuniform motion of the stellar and planetary disks. More
importantly, even with uniform motion the overlap area between the
disks is not a linear function of time [see equation~(1) of Mandel \&
Agol (2002)]. In addition, the bottom of a transit light curve is not
flat because real stellar disks do not have uniform intensity, as
explained in the next section.

\bigskip
\noindent
\textbf{ 2.5 Limb darkening}
\bigskip

\begin{figure}
  \epsscale{1.0}
  \plotone{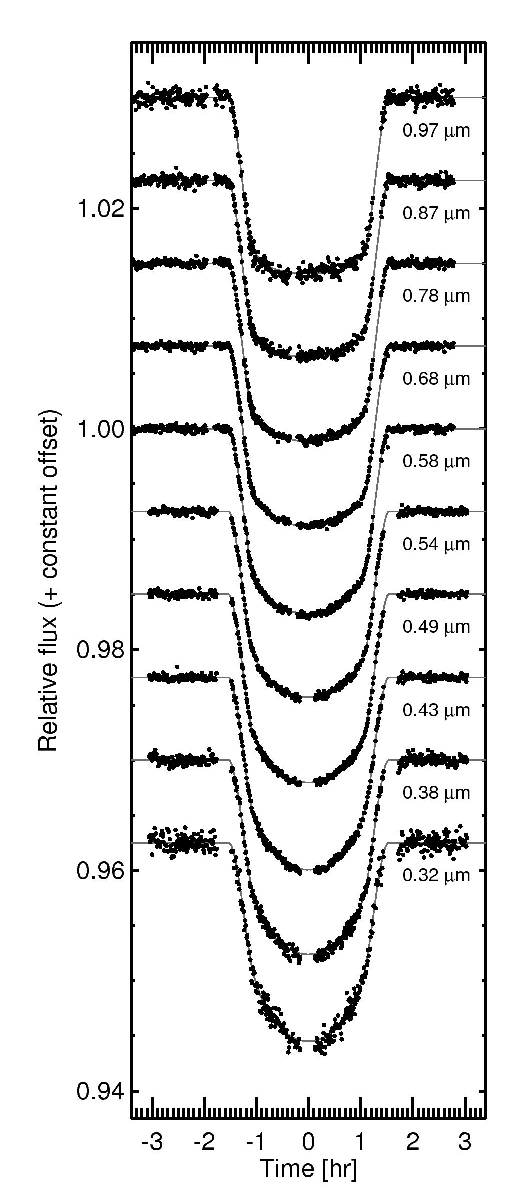}
  \caption{\small Transits of the giant planet HD~209458b observed at
    wavelengths ranging from 0.32~$\mu$m (bottom) to 0.97~$\mu$m
    (top). At shorter wavelengths, the limb darkening of the star is
    more pronounced, and the bottom of the light curve is more
    rounded. The data were collected with the {\it Hubble Space
      Telescope} by Knutson et
    al.~(2007a).\label{fig:knutson-hd209458}}
\end{figure}

Real stellar disks are brighter in the middle and fainter at the edge
(the limb), a phenomenon known as {\it limb darkening}. This causes
the flux decline during a transit to be larger than $k^2$ when the
planet is near the center of the star, and smaller than $k^2$ when the
planet is near the limb. The effect on the light curve is to round off
the bottom and blur the 2nd and 3rd contact points, as shown in
Figure~\ref{fig:knutson-hd209458}. Limb darkening is a
consequence of variations in temperature and opacity with altitude in
the stellar atmosphere. The sight-line to the limb follows a highly
oblique path into the stellar atmosphere, and therefore an optical
depth of unity is reached at a higher altitude, where the temperature
is cooler and the radiation is less intense. The resulting intensity
profile $I(X,Y)$ is often described with a fitting formula such as
\begin{equation}
I \propto 1 - u_1(1-\mu) - u_2(1-\mu)^2,
\label{eq:ld-quadratic}
\end{equation}
where $\mu\equiv\sqrt{1-X^2-Y^2}$ and $\{u_1,u_2\}$ are constant
coefficients that may be calculated from stellar-atmosphere models or
measured from a sufficiently precise transit light curve. The decision
to use the quadratic function of equation~(\ref{eq:ld-quadratic}) or
another of the various limb-darkening ``laws'' (better described as
fitting formulas) is somewhat arbitrary. Claret~(2004) provides a
compilation of theoretical coefficients, and advocates a
four-parameter law. Southworth~(2008) investigates the results of
fitting the same data set with different limb-darkening laws. Mandel
\& Agol (2002) give accurate expressions for $\alpha_{\rm tra}(t)$ for
some limb-darkening laws, and Gim{\'e}nez (2006) shows how to compute
$\alpha_{\rm tra}(t)$ for an arbitrary law based on earlier work by
Kopal (1979).

By using one of these methods to calculate the flux of a limb-darkened
disk with a circular obstruction, it is usually possible to model real
transit light curves to within the measurement errors (see Section
4.3). In principle, calculations of occultation light curves should
take the planetary limb darkening into account, though the precision
of current data has not justified this level of detail. Likewise, in
exceptional cases it may be necessary to allow for departures from
circular shapes, due to rotational or tidal deformation (see Section
6). Modelers of eclipsing binary stars have long needed to take into
account these and other subtle effects (Kallrath \& Milone 2009,
Hilditch 2001).

More generally, the loss of light depends on the intensity of the
particular patch of the photosphere that is hidden by the planet. The
planet provides a raster scan of the stellar intensity across the
transit chord. In this manner, star spots and plages can be detected
through the flux anomalies that are observed when the planet covers
them. (An example is given in the top right panel of
Figure~\ref{fig:lightcurves}.) Even spots that are not along the
transit chord can produce observable effects, by causing variations in
$F_\star$ and thereby causing $\delta_{\rm tra}$ to vary from transit
to transit.

\bigskip
\centerline{\textbf{ 3. SCIENCE FROM ECLIPSES}}
\bigskip
\noindent
\textbf{ 3.1 Determining absolute dimensions}
\bigskip

Chief among the reasons to observe transits is to determine the mass
and radius of the planet. Ideally one would like to know the mass in
kilograms, and the radius in kilometers, to allow for physical
modeling and comparisons with solar system planets. With only a
transit light curve, this is impossible. The light curve by itself
reveals the planet-to-star radius ratio $k \equiv R_p/R_\star \approx
\sqrt{\delta}$ but not the planetary radius, and says nothing about
the planetary mass.

To learn the planetary mass, in addition to the light curve one needs
the radial-velocity orbit of the host star, and in particular the
velocity semiamplitude $K_\star$. Using equation~(66) of the chapter
by Murray and Correia, and Kepler's third law, we may write
\begin{equation}
\frac{M_p}{(M_p + M_\star)^{2/3}} =
\frac{K_\star\sqrt{1-e^2}}{\sin i} \left( \frac{P}{2\pi G} \right)^{1/3}.
\end{equation}
The observation of transits ensures $\sin i\approx 1$, and thereby
breaks the usual $M_p\sin i$ degeneracy. However, the planetary mass
still cannot be determined independently of the stellar mass. In the
usual limit $M_p \ll M_\star$, the data determine $M_p/M_\star^{2/3}$
but not $M_p$ itself.

To determine the absolute dimensions of the planet, one must
supplement the transit photometry and radial-velocity orbit with some
external information about the star. Depending on the star, the
available information may include its luminosity, spectral type, and
spectrally-derived photospheric properties (effective temperature,
surface gravity, and metallicity). A typical approach is to seek
consistency between those data and stellar-evolutionary models, which
relate the observable properties to the stellar mass, radius,
composition, and age. In special cases it may also be possible to pin
down the stellar properties using interferometry (Baines et al.\
2009), asteroseismology (Stello et al.\ 2009), or an eclipsing
companion star.

Besides $\delta$, the transit light curve offers the observables
$T_{\rm tot}$ and $T_{\rm full}$ (or $T$ and $\tau$), which can be
used to solve for the impact parameter $b$ and the {\it scaled stellar
  radius} $R_\star/a$. For non-grazing transits, in the limit $R_p \ll
R_\star \ll a$ we may invert
equations~(\ref{eq:t-tot}-\ref{eq:t-full}) to obtain the approximate
formulas
\begin{eqnarray}
b^2 & = & \frac{(1-\sqrt{\delta})^2 - (T_{\rm full}/T_{\rm tot})^2 (1 + \sqrt{\delta})^2}
{1 - (T_{\rm full}/T_{\rm tot})^2} \\
\frac{R_\star}{a} & = & \frac{\pi}{2 \delta^{1/4}}
\frac{\sqrt{T_{\rm tot}^2 - T_{\rm full}^2}}{P}
  \left( \frac{1+e\sin\omega}{\sqrt{1-e^2}} \right).
\end{eqnarray}
If in addition $\tau \ll T$, such as is the case for small planets on
non-grazing trajectories, the results are simplified still further to
\begin{eqnarray}
b^2 & = & 1 - \sqrt{\delta} \frac{T}{\tau}, \\
\frac{R_\star}{a} & = & \frac{\pi}{\delta^{1/4}}
\frac{\sqrt{T\tau}}{P}
  \left( \frac{1+e\sin\omega}{\sqrt{1-e^2}} \right).
\end{eqnarray}
The orbital inclination $i$ may then be obtained using
equation~(\ref{eq:b-tra}). These approximations are useful for
theoretical calculations and for developing an intuition about how the
system parameters affect the observable light curve. For example,
$R_\star/a$ controls the product of $T$ and $\tau$ while $b$ controls
their ratio. However, as mentioned earlier, for fitting actual data
one needs a realistic limb-darkened model, linked to a Keplerian
orbital model.

The dimensionless ratios $R_\star/a$ and $R_p/a$ are important for
several reasons: (i) They set the scale of tidal interactions between
the star and planet. (ii) $R_p/a$ determines what fraction of the
stellar luminosity impinges on the planet, as discussed in Section
3.4. (iii) $R_\star/a$ can be used to determine a particular
combination of the stellar mean density $\rho_\star$ and planetary
mean density $\rho_p$:
\begin{equation}
\rho_\star + k^3 \rho_p = \frac{3\pi}{GP^2}\left( \frac{a}{R_\star} \right)^3.
\label{eq:stellar-mean-density}
\end{equation}
This can be derived from Kepler's third law (Seager \& Mallen-Ornelas
2003). Since $k^3$ is usually small, the second term on the left side
of equation~(\ref{eq:stellar-mean-density}) is often negligible and
$\rho_\star$ can be determined purely from transit photometry.

This method for estimating $\rho_\star$ has proven to be a useful
diagnostic in photometric transit surveys: a true transit signal
should yield a value of $\rho_\star$ that is consistent with
expectations for a star of the given luminosity and spectral
type. Furthermore, once a precise light curve is available,
$\rho_\star$ is a valuable additional constraint on the stellar
properties.

Interestingly, it is possible to derive the planetary surface gravity
$g_p \equiv GM_p/R_p^2$ independently of the stellar properties:
\begin{equation}
g_p = \frac{2\pi}{P} \frac{\sqrt{1-e^2}~K_\star}{(R_p/a)^2 \sin i}.
\end{equation}
This is derived from equation~(66) of the chapter by Murray and
Correia and Kepler's third law (Southworth et al.~2007).

In short, precise transit photometry and Doppler velocimetry lead to
correspondingly precise values of the stellar mean density
$\rho_\star$ and planetary surface gravity $g_p$. However, the errors
in $M_p$ and $R_p$ are ultimately limited by the uncertainties in the
stellar properties.

\bigskip
\noindent
\textbf{ 3.2 Timing of eclipses}
\bigskip

The orbital period $P$ can be determined by timing a sequence of
transits, or a sequence of occultations, and fitting a linear
function
\begin{equation}
t_c[n] = t_c[0] + nP, \label{eq:linear-ephemeris}
\end{equation}
where $t_c[n]$ is the time of conjunction of the $n$th event. The
times must first be corrected to account for the Earth's orbital
motion and consequent variations in the light travel time. When
comparing transit and occultation times, one must further correct for
the light travel time across the line-of-sight dimension of the
planetary orbit. As long as there is no ambiguity in $n$, the error in
$P$ varies inversely as the total number of eclipses spanned by the
observations, making it possible to achieve extraordinary precision in
$P$.

If the orbit does not follow a fixed ellipse---due to forces from
additional bodies, tidal or rotational bulges, general relativity, or
other non-Keplerian effects---then there will be variations in the
interval between successive transits, as well as the interval between
transits and occultations and the shape of the transit light
curve. These variations may be gradual parameter changes due to
precession (Miralda-Escud\'e 2002), or short-term variations due to
other planets (Holman \& Murray 2005, Agol et al.~2005) or moons
(Kipping 2009). The effects can be especially large for bodies in
resonant orbits with the transiting planet. By monitoring transits one
might hope to detect such bodies, as discussed in the chapter by
Fabrycky.

When transits and occultations are both seen, a powerful constraint on
the shape of the orbit is available. For a circular orbit, those
events are separated in time by $P/2$, but more generally the time
interval depends on $e$ and $\omega$. To first order in $e$,
integrating $dt/df$ between conjunctions gives
\begin{equation}
\Delta t_c \approx \frac{P}{2}\left[1 + \frac{4}{\pi} e\cos\omega\right].
\label{eq:delta-tc-tra-occ}
\end{equation}
In this case, the timing of transits and occultations gives an
estimate of $e\cos\omega$. Likewise the relative durations of
the transit and the occultation depend on the complementary parameter
$e\sin\omega$,
\begin{equation}
\frac{T_{\rm occ}}{T_{\rm tra}} \approx \frac{1 + e\sin\omega}{1 - e\sin\omega}.
\end{equation}
Sterne~(1940) and de Kort (1954) give the lengthy exact results for
arbitrary $e$ and $i$. Because the uncertainty in $\Delta t_c/P$ is
typically smaller than that in $T_{\rm tra}/T_{\rm occ}$ (by a factor
of $P/T$), the eclipse data constrain $e\cos\omega$ more powerfully
than $e\sin\omega$.

The resulting bounds on $e$ are often valuable. For example, planets
on close-in eccentric orbits are internally heated by the friction
that accompanies the time-variable tidal distortion of the
planet. Empirical constraints on $e$ thereby help to understand the
thermal structure of close-in planets. For more distant planets,
bounds on $e$ are helpful in the statistical analysis of exoplanetary
orbits. As described in the chapter by Cumming and in Part~V of this
volume, the observed eccentricity distribution of planetary orbits is
a clue about the processes of planet formation and subsequent orbital
evolution.

Eclipse-based measurements of $t_{\rm tra}$, $t_{\rm occ}$, and $P$
are almost always more precise than those based on spectroscopic or
astrometric orbital data. The eclipse-based results can greatly
enhance the analysis of those other data. For example, the usual
radial-velocity curve has 6 parameters, but if $t_{\rm tra}$, $t_{\rm
  occ}$, and $P$ are known from eclipses, the number of free
parameters is effectively reduced to 3, thereby boosting the
achievable precision in the other 3 parameters.

\bigskip
\noindent
\textbf{ 3.3 Transmission spectroscopy}
\bigskip

We have been implicitly assuming that the planetary silhouette has a
sharp edge, but in reality the edge is fuzzy. For gas giant planets
there is no well-defined surface, and even planets with solid surfaces
may have thick atmospheres. During a transit, a small portion of the
starlight will be filtered through the upper atmosphere of the planet,
where it is only partially absorbed. The absorption will be
wavelength-dependent due to the scattering properties of atoms and
molecules in the planetary atmosphere. At the wavelength of a strong
atomic or molecular transition, the atmosphere is more opaque, and the
planet's effective silhouette is larger. This raises the prospect of
measuring the {\it transmission spectrum} of the planet's upper
atmosphere and thereby gaining knowledge of its composition.

To calculate the expected signal one must follow the radiative
transfer of the incident starlight along a grazing trajectory through
the planet's stratified atmosphere. The calculation is rather
complicated (see, e.g., Seager \& Sasselov 2000, Brown 2001) but the
order of magnitude of the effect is easily appreciated. For a strong
transition, the effective size of the planet grows by a few
atmospheric scale heights $H$, where
\begin{equation}
H = \frac{k_BT}{\mu_m g}
\end{equation}
and $T$ is the temperature, $\mu_m$ is the mean molecular mass, $g$ is
the local gravitational acceleration, and $k_B$ is Boltzmann's
constant.  Defining $R_p$ as the radius within which the planet is
optically thick at all wavelengths, the extra absorption due to the
optically thin portion of the atmosphere causes the transit depth to
increase by
\begin{equation}
\Delta \delta = \frac{\pi(R_p + N_H H)^2}{\pi R_\star^2} -
\frac{\pi R_p^2}{\pi R_\star^2}
 \approx 2 N_H \delta \left(\frac{H}{R_p}\right),
\label{eq:transmission-fom}
\end{equation}
where $N_H$, the number of scale heights, is of order unity. The
signal is most readily detectable for planets with large $H$: low
surface gravity, low mean molecular mass, and high temperature. For a
``hot Jupiter'' around a Sun-like star ($\delta = 0.01$, $T\approx
1300$~K, $g\approx 25$~m~s$^{-2}$, $\mu_m = 2$~amu) the signal is
$\Delta\delta \sim 10^{-4}$. For an Earth-like planet around a
Sun-like star ($\delta = 10^{-4}$, $T\approx 273$~K, $g\approx
10$~m~s$^{-2}$, $\mu_m = 28$~amu), the signal is $\Delta\delta \sim
10^{-6}$.

The signal can be detected by observing a transit light curve at
multiple wavelengths, using different filters or a spectrograph. One
then fits a limb-darkened light-curve model to the time series
obtained at each wavelength, requiring agreement in the orbital
parameters and allowing a value of $\delta$ specific to each
wavelength. The resulting variations in $\delta(\lambda)$ are expected
to be of the order of magnitude given by
equation~(\ref{eq:transmission-fom}). It is best to gather all of the
data at the same time, because intrinsic stellar variability is also
chromatic.

\bigskip
\noindent
\textbf{ 3.4 Occultation spectroscopy}
\bigskip

As discussed in Section 2.4, when the planet is completely hidden the
starlight declines by a fraction $\delta_{\rm occ} = k^2 I_p/I_\star$,
where $k$ is the planet-to-star radius ratio and $I_p/I_\star$ is the
ratio of disk-averaged intensities. Observations spanning occultations
thereby reveal the relative brightness of the planetary disk, if $k$
is already known from transit observations. The planetary radiation
arises from two sources: thermal radiation and reflected
starlight. Because the planet is colder than the star, the thermal
component emerges at longer wavelengths than the reflected component.

For the moment we suppose that the planet is of uniform brightness,
and that the observing wavelength is long enough for thermal emission
to dominate. Approximating the planet and star as blackbody radiators,
\begin{equation}
\delta_{\rm occ}(\lambda) = k^2~\frac{B_\lambda(T_p)}{B_\lambda(T_\star)}~
\longrightarrow~
k^2~\frac{T_p}{T_\star}
\label{eq:occultation-depth}
\end{equation}
where $B_\lambda(T)$ is the Planck function,
\begin{equation}
B_\lambda(T) \equiv \frac{2hc^2}{\lambda^5} \frac{1}{e^{hc/(\lambda k_B T)} - 1}~
\longrightarrow~
\frac{2ck_BT}{\lambda^4},
\end{equation}
in which $T$ is the temperature, $\lambda$ is the wavelength, $h$ is
Planck's constant, and $c$ is the speed of light. The limiting cases
are for the ``Rayleigh-Jeans'' limit $\lambda \gg hc/(k_B T)$. The
decrement $\delta_{\rm occ}$ that is observed by a given instrument is
obtained by integrating equation~(\ref{eq:occultation-depth}) over the
bandpass.

Even when the planetary radiation is not described by the Planck law,
one may define a {\it brightness temperature}\, $T_b(\lambda)$ as the
equivalent blackbody temperature that would lead to the observed value
of $\delta_{\rm occ}(\lambda)$. The brightness temperature is
sometimes a convenient way to describe the wavelength-dependent
intensity even when it is not thermal in origin.

There may be departures from a blackbody spectrum---spectral
features---discernible in the variation of brightness temperature with
wavelength. In contrast with transmission spectroscopy, which refers
to starlight that grazes the planetary limb (terminator), here we are
referring to the emission spectrum of the planet averaged over the
visible disk of the dayside. Occultation spectroscopy and transit
spectroscopy thereby provide different and complementary information
about the planetary atmosphere.

It is possible to measure the reflectance spectrum of the planet's
dayside by observing at shorter wavelengths, or accurately subtracting
the thermal emission. The occultation depth due to reflected light
alone is
\begin{equation}
\delta_{\rm occ}(\lambda) = A_\lambda\left(\frac{R_p}{a}\right)^2,
\end{equation}
where $A_\lambda$ is the {\it geometric albedo}, defined as the flux
reflected by the planet when viewed at opposition (full phase),
divided by the flux that would be reflected by a flat and perfectly
diffusing surface with the same cross-sectional area as the
planet. One of the greatest uncertainties in atmospheric modeling is
the existence, prevalence, and composition of clouds.  Since clouds
can produce very large albedo variations, reflectance spectroscopy may
help to understand the role of clouds in exoplanetary atmospheres.

For a close-in giant planet, the reflectance signal is $\sim$$10^{-4}$
while for an Earthlike planet at 1~AU it is $\sim$$10^{-9}$. The
detection prospects are better for closer-in planets. However, the
closest-in planets are also the hottest, and their radiation may be
dominated by thermal emission. Another consideration is that planets
with $T>1500$~K are expected to be so hot that all potentially
cloud-forming condensable materials are in gaseous form. The
theoretically predicted albedos are very low, of order $10^{-3}$, due
to strong absorption by neutral sodium and potassium.

Real planets do not have uniformly bright disks. Gaseous planets are
limb darkened or brightened, and may have latitudinal zones with high
contrast, like Jupiter. Rocky planets may have surface features and
oceans. To the extent that departures from uniform brightness could be
detected, occultation data would provide information on the
spatially-resolved planetary dayside. Specifically, the light curve of
the ingress or egress of an occultation gives a one-dimensional
cumulative brightness distribution of the planet.

\bigskip
\noindent
\textbf{ 3.5 The Rossiter-McLaughlin effect}
\bigskip

In addition to the spectral variations induced by the planetary
atmosphere, there are spectral variations arising from the spatial
variation of the stellar spectrum across the stellar disk. The most
pronounced of these effects is due to stellar rotation: light from the
approaching half of the stellar disk is blueshifted, and light from
the receding half is redshifted. Outside of transits, rotation
broadens the spectral lines but does not produce an overall Doppler
shift in the disk-integrated starlight. However, when the planet
covers part of the blueshifted half of the stellar disk, the
integrated starlight appears slightly redshifted, and vice versa.

\begin{figure*}
  \epsscale{1.25}
  \plotone{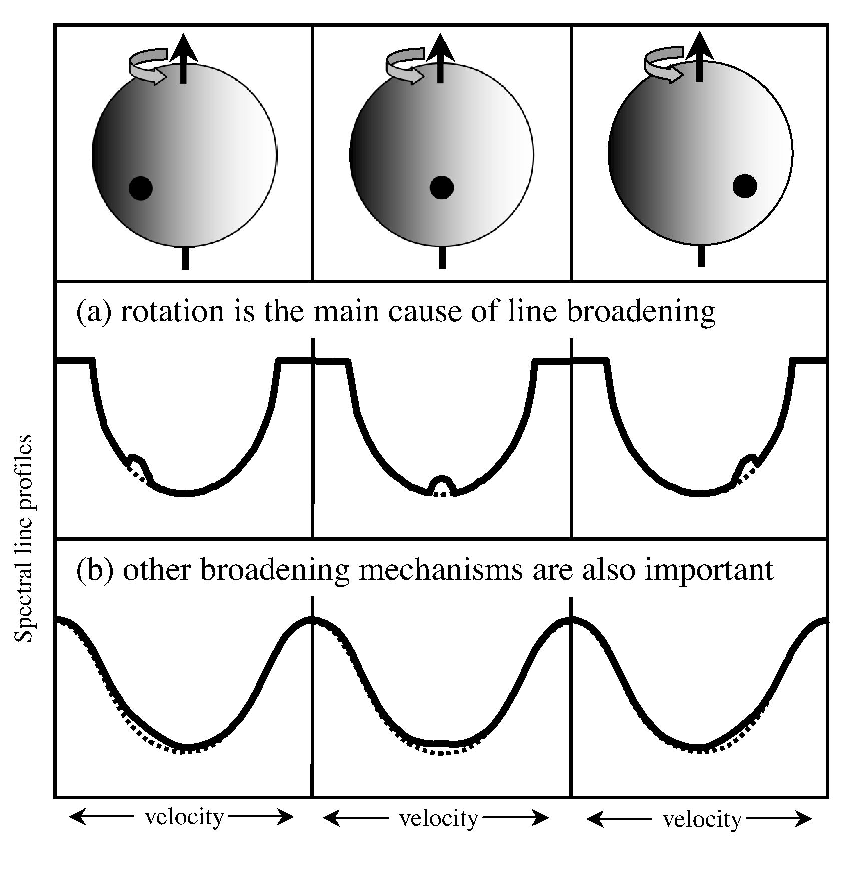}
  \caption{\small Illustration of the Rossiter-McLaughlin (RM) effect.
    The three columns show three successive phases of a transit. The
    first row shows the stellar disk, with the grayscale representing
    the projected rotation velocity: the approaching limb is black and
    the receding limb is white. The second row shows the corresponding
    stellar absorption line profiles, assuming rotation to be the
    dominant broadening mechanism. The ``bump'' occurs because the
    planet hides a fraction of the light that contributes a particular
    velocity to the line-broadening kernel. The third row shows the
    case for which other line-broadening mechanisms are important;
    here the RM effect is manifested only as an ``anomalous Doppler
    shift.'' Adapted from Gaudi \& Winn (2007).\label{fig:rmlines}}
\end{figure*}

Thus, the transit produces a time-variable spectral distortion that is
usually manifested as an ``anomalous'' radial velocity, i.e., a
Doppler shift that is greater or smaller than the shift expected from
only the star's orbital motion. Figure~\ref{fig:rmlines}
illustrates this effect. It is known as the Rossiter-McLaughlin (RM)
effect, after the two astronomers who made the first definitive
observations of this kind for binary stars, in 1924.

The maximum amplitude of the anomalous radial velocity is
approximately
\begin{equation}
  \Delta V_{\rm RM} \approx k^2 \sqrt{1-b^2} (v_\star \sin i_\star),
\end{equation}
where $v_\star \sin i_\star$ is the line-of-sight component of the
stellar equatorial rotation velocity. For a Sun-like star
($v_\star\sin i_\star$=2~km~s$^{-1}$), the maximum amplitude is
$\sim$20~m~s$^{-1}$ for a Jovian planet and $\sim$0.2~m~s$^{-1}$ for a
terrestrial planet. This amplitude may be comparable to (or even
larger than) the amplitude of the spectroscopic orbit of the host
star. Furthermore it is easier to maintain the stability of a
spectrograph over the single night of a transit than the longer
duration of the orbital period. Hence the RM effect is an effective
means of detecting and confirming transits, providing an alternative
to photometric detection.

\begin{figure*}
  \epsscale{1.75}
  \plotone{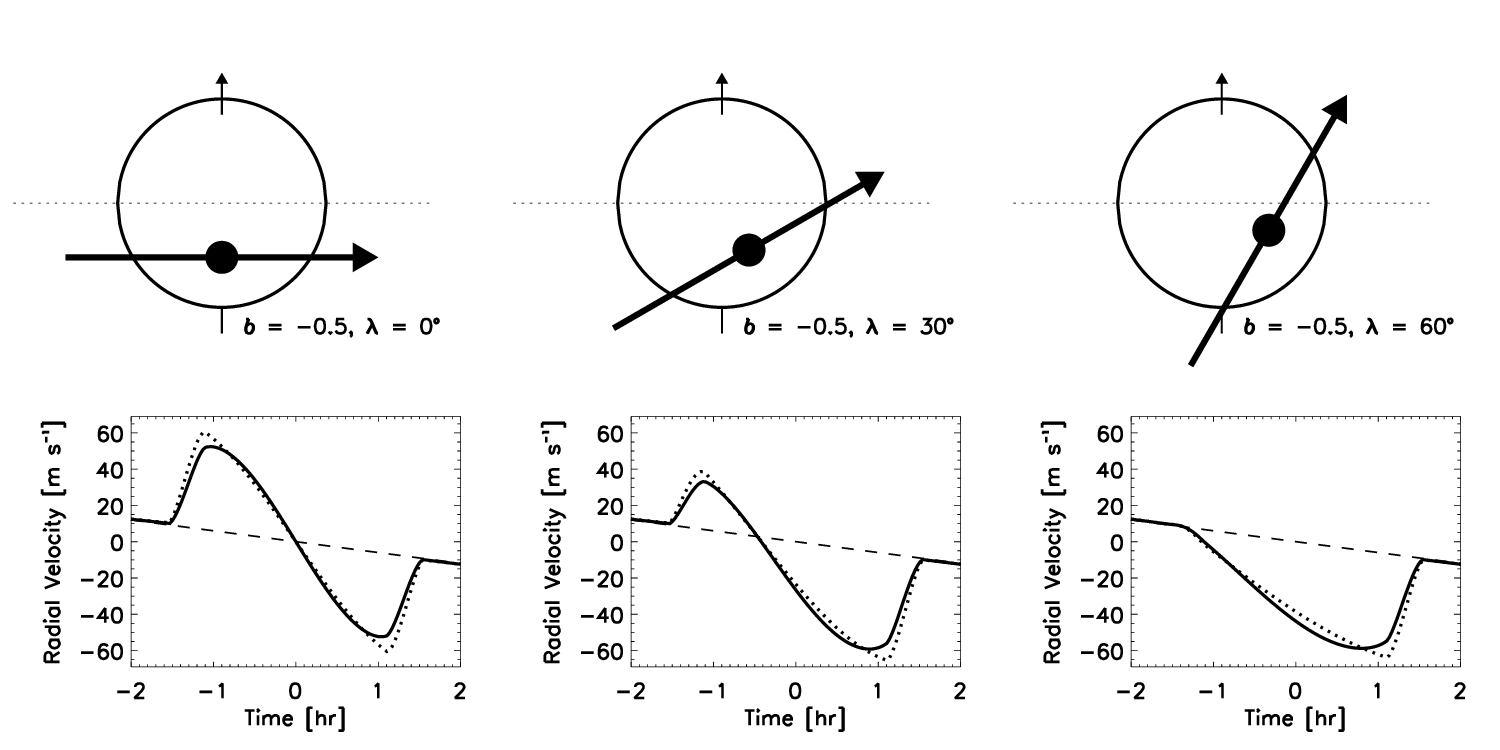}
  \caption{\small Using the RM effect to measure the angle $\lambda$
    between the sky projections of the orbital and stellar-rotational
    axes. Three different possible trajectories of a transiting planet
    are shown, along with the corresponding RM signal. The
    trajectories all have the same impact parameter and produce the
    same light curve, but they differ in $\lambda$ and produce
    different RM curves. The dotted lines are for the case of no limb
    darkening, and the solid lines include limb darkening. From Gaudi
    \& Winn (2007).\label{fig:trajectories}}
\end{figure*}

In addition, by monitoring the anomalous Doppler shift throughout a
transit, it is possible to measure the angle on the sky between the
planetary orbital axis and the stellar rotation
axis. Figure~\ref{fig:trajectories} shows three trajectories of a
transiting planet that have the same impact parameter, and hence
produce identical light curves, but that have different orientations
relative to the stellar spin axis, and hence produce different RM
signals. The signal for a well-aligned planet is antisymmetric about
the midtransit time (left panels), whereas a strongly misaligned
planet that blocks only the receding half of the star will produce
only an anomalous blueshift (right panels).

A limitation of this technique is that it is only sensitive to the
angle between the {\it sky projections} of the spin and orbital
angular momentum vectors. The true angle between those vectors is
usually poorly constrained because $i_\star$ is unknown. Nevertheless
it may be possible to tell whether the planetary orbit is prograde or
retrograde, with respect to the direction of stellar rotation. It is
also possible to combine results from different systems to gain
statistical knowledge about spin-orbit alignment.

More broadly, just as transit photometry provides a raster scan of the
intensity of the stellar photosphere along the transit chord, RM data
provide a raster scan of the line-of-sight velocity field of the
photosphere. This gives an independent measure of the projected
rotation rate $v_\star\sin i_\star$, and reveals the velocity
structure of starspots or other features that may exist on the
photosphere.

\bigskip
\bigskip
\centerline{\textbf{ 4. OBSERVING ECLIPSES}}
\bigskip

\noindent
\textbf{ 4.1 Discovering eclipsing systems}
\bigskip

\begin{figure*}
  \epsscale{1.5}
  \plotone{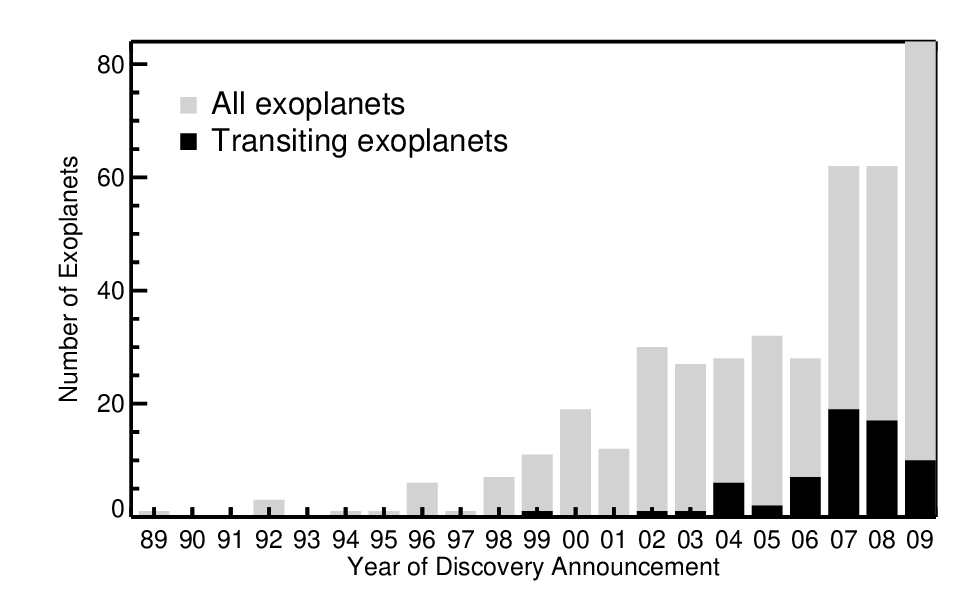}
  \caption{\small Rate of exoplanet discovery. The light bars show the
    number of announcements of newly discovered planets, and the dark
    bars show the subset of transiting planets. Compiled with data
    from exoplanet.eu, an encyclopedic web site maintained by
    J.~Schneider.\label{fig:disc-hist}}
\end{figure*}

Figure~\ref{fig:disc-hist} shows the rate of exoplanet discoveries
over the last 20 years, highlighting the subset of planets known to
transit. Transits have been discovered in two ways. One way is to find
the planet in a Doppler survey (see chapter by Lovis and Fischer), and
then check for transits by monitoring the brightness of the star
throughout the inferior conjunction. This was the path to discovery
for 6 of the 64 known systems, including the first example, HD~209458b
(Charbonneau et al.~2000, Henry et al.~2000, Mazeh et al.~2000). The
probability that transits will occur is given by equation~(9) and the
times of inferior conjunction can be calculated from the parameters of
the spectroscopic orbit (Kane 2007).

The other way is to conduct photometric surveillance of stars that are
not yet known to have any planets, an idea dating back to
Struve~(1952). As an illustration let us consider the requirements for
a program to discover hot Jupiters. As discussed in Section 2.2, a
bare minimum of $10^3$ Sun-like stars must be examined to find a
transiting planet with an orbital distance of $0.05$~AU. The
observations must be precise enough to detect a 1\% flux drop, and
must extend for at least several times longer than the 3~day orbital
period. The difficulty increases dramatically for smaller planets in
wider orbits: in an idealized survey of nearby field stars limited by
photon noise (see Section 4.2) the sensitivity to transiting planets
scales approximately as $R_p^6/P^{5/3}$ (Gaudi 2005; Gaudi, Seager, \&
Mallen-Ornelas 2005). More realistic calculations take into account
the spread in radius and luminosity among the surveyed stars, the
duration and time-sampling of observations, the appropriate threshold
value of the signal-to-noise ratio, and other factors (see, e.g.,
Pepper et al.\ 2003, Beatty \& Gaudi 2008).

Photometric surveys are much less efficient than Doppler surveys, in
the sense that only a small fraction of planets transit, and even when
transits occur they are underway only a small fraction of the time. On
the other hand, the starting equipment for a photometric survey is
modest, consisting in some cases of amateur-grade telescopes or
telephoto lenses and cameras, whereas Doppler surveys require upfront
a large telescope and sophisticated spectrograph. For these reasons
the royal road of eclipses enticed many astronomers to embark on
photometric transit surveys. More than a dozen surveys were
undertaken, including a few longitudinally-distributed networks to
provide more continuous time coverage. Major efforts were made to
automate the observations, and to develop algorithms for precise
wide-field photometry (Tamuz et al.~2005, Kov{\'a}cs et al.~2005) and
transit detection (Kov{\'a}cs et al.~2002). Horne~(2003) took stock of
all the ongoing and planned surveys, and ventured to predict a bounty
of 10--100 new planets per month, in an article subtitled ``Hot
Jupiters Galore.''

The royal road turned out to have some potholes. One obstacle was the
high rate of ``false positives,'' signals that resemble planetary
transits but are actually grazing eclipses of a binary star, or an
unresolved combination of an eclipsing binary star and a third
star. In the latter case, the deep eclipses of the binary star are
diluted to planet-like proportions by the constant light of the third
star. In some surveys the false positives outnumbered the planets by
10 to 1. Ruling them out required spectroscopy with large telescopes,
which became the bottleneck in the discovery process (see, e.g.,
O'Donovan et al.~2007). Another obstacle was correlated noise (``red
noise'') in the survey photometry (see Section 4.3). Pont et
al.~(2006) showed that red noise slashed the sensitivity of the search
algorithms, thereby providing a quantitative solution to the ``Horne
problem'' of why transiting planets were not being found as rapidly as
expected.

Only a few of the surveyors were able to overcome these obstacles. The
5 most successful surveys were OGLE, which used a 1~m telescope to survey
14-16th magnitude stars; and the TrES, XO, HAT, and SuperWASP surveys,
which used $\approx$0.1~m lenses to survey 10-12th magnitude
stars. Many of today's transiting planets are named after these
surveys: for more details see Udalski et al.~(2002), Alonso et
al.~(2004), McCullough et al.~(2005), Bakos et al.~(2007), and
Pollacco et al.~(2006). The OGLE planets were found first, although
the brighter stars from the wider-field surveys were far more amenable
to false-positive rejection and detailed characterization.

On the other end of the cost spectrum are photometric surveys
performed from space. By avoiding the deleterious effects of the
Earth's atmosphere, it is possible to beat the precision of
ground-based observations (see Section 4.2). It is also possible to
avoid the usual interruptions due to the vagaries of the weather, and
even to avoid the day-night cycle if the spacecraft is in an orbit far
from Earth. The two ongoing space-based missions {\it CoRoT} and {\it
  Kepler} are described in Sections 5.1 and 6, respectively.

One space-based survey that did not discover any planets neverthless
produced an interesting result. Gilliland et al.~(2000) used the {\it
  Hubble Space Telescope} to seek close-in giant planets in the
globular cluster 47~Tucanae. No transits were found, even though 17
were expected based on the survey characteristics and the observed
frequency of close-in giant planets around nearby field stars. The
absence of close-in planets could be due to the crowded stellar
environment (which could inhibit planet formation or disrupt planetary
systems) or the cluster's low metallicity (which has been found to be
correlated with fewer planets in the Doppler surveys; see the chapters
by Lovis and Fischer and by Cumming). The null results of a subsequent
survey of the less-crowded outskirts of 47~Tucanae suggest that
crowding is not the primary issue (Weldrake et al.~2005).

\bigskip
\noindent
\textbf{ 4.2 Measuring the photometric signal}
\bigskip

Even after a transiting planet is discovered, it takes a careful hand
to measure the transit signal precisely enough to achieve the
scientific goals set forth in Section 3. The loss of light is only 1\%
for a Sun-like star crossed by a Jupiter-sized planet, and 0.01\% for
an Earth-sized planet. Occultations produce still smaller
signals. This is the domain of precise time-series differential
photometry. The term ``differential'' applies because only the
fractional variations are of interest, as opposed to the actual
intensity in Janskys or other standardized units. Eclipses can also be
detected via spectroscopy (Section~3.5) or polarimetry (Carciofi \&
Magalh\~aes 2005) but here we focus on the photometric signal, with
emphasis on ground-based optical observations.

First you must know when to observe. To observe an eclipse requires a
triple coincidence: the eclipse must be happening, the star must be
above the horizon, and the Sun must be down. Transit times can be
predicted based on a sequence of previously measured transit times, by
fitting and extrapolating a straight line
(equation~\ref{eq:linear-ephemeris}). Occultation times can also be
predicted from a listing of transit times, but are subject to
additional uncertainty due to the dependence on $e$ and $\omega$
(equation~\ref{eq:delta-tc-tra-occ}).

Next you should monitor the flux of the target star along with other
nearby stars of comparable brightness. The measured fluxes are
affected by short-term variations in atmospheric transparency, as well
as the gradual change in the effective atmospheric path length (the
{\it airmass}) as the star rises and sets. However, the ratios of the
fluxes between nearby stars are less affected. As long as some of the
comparison stars are constant in brightness then the relative flux of
the target star can be tracked precisely.

This task is usually accomplished with a charge-coupled device (CCD)
imaging camera and software for calibrating the images, estimating the
sky background level, and counting the photons received from each star
in excess of the sky level ({\it aperture photometry}). Howell~(1999)
explains the basic principles of CCDs and aperture photometry. For
more details on differential aperture photometry using an ensemble of
comparison stars, see Gilliland \& Brown~(1988), Kjeldsen \&
Frandsen~(1992), and Everett \& Howell~(2001). In short, the fluxes of
the comparison stars are combined, and the flux of the target star is
divided by the comparison signal, giving a time series of relative
flux measurements spanning the eclipse with as little noise as
possible. For bright stars observed at optical wavelengths, among the
important noise sources are photon noise, scintillation noise,
differential extinction, and flat-fielding errors, which we now
discuss in turn.

{\it Photon noise} refers to the unavoidable fluctuations in the
signal due to the quantization of light. It is also called {\it
  Poisson noise} because the photon count rate obeys a Poisson
distribution. If a star delivers $N$ photons~s$^{-1}$ on average, then
the standard deviation in the relative flux due to photon noise is
approximately $(N\Delta t)^{-1/2}$ for an exposure lasting $\Delta t$
seconds. This noise source affects the target and comparison stars
independently. The sky background also introduces Poisson noise, which
can be troublesome for faint stars or infrared wavelengths. The photon
noise in the comparison signal can be reduced by using many bright
comparison stars. Beyond that, improvement is possible only by
collecting more photons, using a bigger telescope, a more efficient
detector, or a wider bandpass.

{\it Scintillation} is caused by fluctuations in the index of
refraction of air. The more familiar term is ``twinkling.'' For
integration times $\Delta t \gsim 1$~s, the standard deviation in the
relative flux due to scintillation is expected to scale with telescope
diameter $D$, observatory altitude $h$, integration time, and airmass
as
\begin{equation}
\sigma_{\rm scin} = \sigma_0 \frac{({\rm Airmass})^{7/4}}{D^{2/3} (\Delta t)^{1/2}}
\exp\left(-\frac{h}{{\rm 8000~m}}\right),
\end{equation}
based on a theory of atmospheric turbulence by Reiger~(1963), with
empirical support from Young (1967) and others. The coefficient
$\sigma_0$ is often taken to be 0.064 when $D$ is expressed in
centimeters and $\Delta t$ in seconds, but this must be understood to
be approximate and dependent on the local meteorology. Scintillation
affects both the target and comparison stars, although for
closely-spaced stars the variations are correlated (Ryan \& Sandler
1998). One also expects scintillation noise to decrease with
wavelength. Thus, scintillation noise is reduced by employing a large
telescope (or combining results from multiple small telescopes),
choosing nearby comparison stars, and observing at a long wavelength
from a good site.

{\it Differential extinction} is used here as a shorthand for
``second-order color-dependent differential extinction.'' To first
order, if two stars are observed simultaneously, their fluxes are
attenuated by the Earth's atmosphere by the same factor and the flux
ratio is preserved. To second order, the bluer star is attenuated
more, because scattering and absorption are more important at shorter
wavelengths. This effect causes the flux ratio to vary with airmass as
well as with short-term transparency fluctuations. It can be reduced
by choosing comparison stars bracketing the target star in color, or
using a narrow bandpass. The advantage of a narrow bandpass must be
weighed against the increased photon noise.

{\it Flat fielding} is the attempt to correct for nonuniform
illumination of the detector and pixel-to-pixel sensitivity
variations, usually by dividing the images by a calibration image of a
uniformly-lit field. If the stars were kept on the same pixels
throughout an observation, then flat-fielding errors would not affect
the flux ratios. In reality the light from a given star is detected on
different pixels at different times, due to pointing errors, focus
variations, and seeing variations. Imperfect flat fielding coupled
with these variations produce noise in the light curve. The impact of
flat-fielding errors is reduced by ensuring the calibration images
have negligible photon noise, maintaining a consistent pointing, and
defocusing the telescope. Defocusing averages down the interpixel
variations, reduces the impact of seeing variations, and allows for
longer exposure times without saturation, thereby increasing the
fraction of the time spent collecting photons as opposed to resetting
the detector. Defocusing is good for what ails you, as long as the
stars do not blend together.

This discussion of noise sources is not exhaustive; it is in the
nature of noise that no such listing can be complete. For example, the
gain of the detector may drift with temperature, or scattered
moonlight may complicate background subtraction. A general principle
is to strive to keep everything about the equipment and the images as
consistent as possible. Another good practice is to spend at least as
much time observing the star before and after the eclipse as during
the eclipse, to establish the baseline signal and to characterize the
noise.

It is often advisable to use a long-wavelength bandpass, not only to
minimize scintillation and differential extinction, but also to reduce
the effects of stellar limb darkening on the transit light curve. The
degree of limb darkening diminishes with wavelength because the ratio
between blackbody spectra of different temperatures is a decreasing
function of wavelength. Transit light curves observed at longer
wavelengths are ``boxier,'' with sharper corners and flatter
bottoms. All other things being equal, this reduces the statistical
uncertainties in the transit parameters, but other factors should also
be considered. For example, at infrared wavelengths, limb darkening
may be small, but the sky background is bright and variable.

The space-based observer need not worry about scintillation and
differential extinction, and enjoys a low background level even at
infrared wavelengths. Extremely precise photometry is possible,
limited only by the size of the telescope and the degree to which the
detector is well-calibrated. With the {\it Hubble Space Telescope}\, a
precision of approximately $10^{-4}$ per minute-long integration has
been achieved, a few times better than the best ground-based light
curves. Figure~\ref{fig:lightcurves} allows for some side-by-side
comparisons of ground-based and space-based data. However, going to
space is not a panacea. When unforeseen calibration issues arise after
launch, they are difficult to resolve. In low Earth orbit, there are
also unhelpful interruptions due to occultations by the Earth, as well
as problems with scattered Earthshine and cosmic rays. More remote
orbits offer superior observing conditions, but require greater effort
and expense to reach.

\begin{figure}
  \epsscale{1.0}
  \plotone{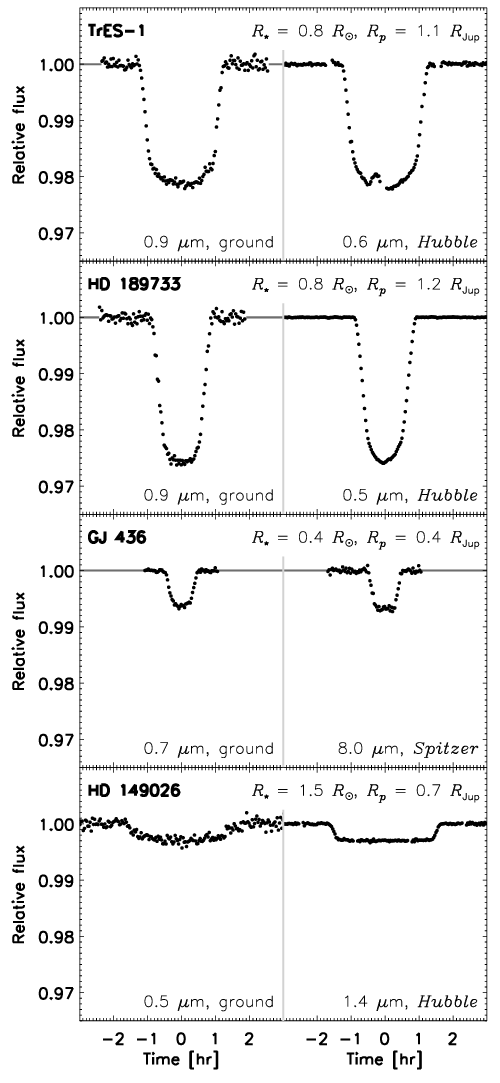}
  \caption{\small Examples of transit light curves based on
    ground-based observations (left) and space-based observations
    (right). In the top right panel, the ``bump'' observed just before
    midtransit is interpreted as the covering of a dark starspot by
    the planet. From upper left to lower right, the references are
    Winn et al.~(2007a), Rabus et al.~(2009), Winn et al.~(2007b),
    Pont et al.\ (2008), Holman et al.\ in prep., Gillon et
    al.~(2007), Winn et al.~(2008), and Carter et
    al.~(2009). \label{fig:lightcurves}}
\end{figure}

\bigskip
\noindent
\textbf{ 4.3 Interpreting the photometric signal}
\bigskip

Once you have an eclipse light curve, the task remains to derive the
basic parameters $\{\delta, T, \tau, t_c\}$ and their uncertainties,
as well as the results for other quantities such as $g_p$ and
$\rho_\star$ that can be derived from those parameters. The analytic
equations given in Section 2 for the eclipse duration, timing, and
other properties are rarely used to analyze data. Rather, a parametric
model is fitted to the data, based on the numerical integration of
Kepler's equation to calculate the relative positions of the star and
planet, as well as one of the prescriptions mentioned in Section 2.4
for computing the loss of light from the limb-darkened stellar
photosphere.

The first task is writing a code that calculates the light curve of
the star-planet system as a function of the orbital parameters and
eclipse parameters. Then, this code is used in conjunction with one of
many standard routines to optimize the parameter values, typically by
minimizing the sum-of-squares statistic,
\begin{equation}
\chi^2 = \sum_{i=1}^N
\left[\frac{f_i({\rm obs}) - f_i({\rm calc})}{\sigma_i}\right]^2,
\label{eq:chisquared}
\end{equation}
where $f_i({\rm obs})$ is the observed value of relative flux at time
$t_i$, $f_i({\rm calc})$ is the calculated flux (depending on the
model parameters), and $\sigma_i$ is the measurement uncertainty. The
techniques found in standard works such as Numerical Recipes (Press et
al.~2007) are applicable here, although a few points of elaboration
are warranted for the specific context of eclipse photometry.

It is common to adopt a Bayesian attitude, in which the parameters are
viewed as random variables whose probability distributions
(``posteriors'') are constrained by the data. This can be done in a
convenient and elegant fashion using the Monte Carlo Markov Chain
(MCMC) method, in which a chain of points is created in parameter
space using a few simple rules that ensure the collection of points
will converge toward the desired posterior. This method gives the full
multidimensional joint probability distribution for all the
parameters, rather than merely giving individual error bars, making it
easy to visualize any fitting degeneracies and to compute posteriors
for any combination of parameters that may be of interest. Although a
complete MCMC briefing is beyond the scope of this chapter, the
interested reader should consult the textbook by Gregory (2005) as
well as case studies such as Holman et al.~(2006), Collier Cameron et
al.~(2007), and Burke et al.~(2007).

A vexing problem is the presence of correlated noise, as mentioned in
Section 4.2. The use of equation~(\ref{eq:chisquared}) is based on the
premise that the measurement errors are statistically independent. In
many cases this is plainly false. Real light curves have bumps,
wiggles, and slopes spanning many data points. These can be attributed
to differential extinction, flat-fielding errors, or astrophysical
effects such as starspots. Thus the number of truly independent
samples is smaller than the number of data points, and the power to
constrain model parameters is correspondingly reduced. Ignoring the
correlations leads to false precision in the derived parameters, but
accounting for correlations is not straightforward and can be
computationally intensive. Some suggestions are given by Pont et
al.~(2006) and Carter \& Winn~(2009a).

The treatment of stellar limb darkening presents another unwelcome
opportunity to underestimate the parameter uncertainties. It is
tempting to adopt one of the standard limb-darkening laws (see Section
2.5) and hold the coefficients fixed at values deemed appropriate for
the host star, based on stellar-atmosphere models. However, with
precise data it is preferable to fit for the coefficients, or at least
to allow for some uncertainty in the atmospheric models. In those few
cases where the data have been precise enough to test the models, the
models have missed the mark (Claret~2009).

Although fitting data is a job for a computer, it is nevertheless
useful to have analytic formulas for the achievable precision in the
eclipse parameters. The formulas are handy for planning observations,
and for order-of-magnitude estimates of the observability (or not) of
effects such as variations in transit times and durations. The
formulas given here are based on the assumptions that the data have
uniform time sampling $\Delta t$, and independent Gaussian errors
$\sigma$ in the relative flux. A useful figure of merit is $Q\equiv
\sqrt{N}\delta/\sigma$, where $\delta$ is the transit depth and $N$ is
the number of data points obtained during the transit. A Fisher
information analysis (essentially a glorified error propagation) leads
to estimates for the 1$\sigma$ uncertainties in the transit parameters
(Carter et al.~2008):
\begin{eqnarray}
\sigma_{\delta} & \approx & Q^{-1} \delta,\\
\sigma_{t_c} & \approx & Q^{-1} T\sqrt{\tau/2T}, \\
\sigma_{T} & \approx & Q^{-1} T\sqrt{2\tau/T}, \\
\sigma_{\tau} & \approx & Q^{-1} T\sqrt{6\tau/T},
\end{eqnarray}
which are valid when $\delta\ll 1$, limb darkening is weak, and the
out-of-transit flux is known precisely. In this case, $\sigma_{t_c} <
\sigma_T < \sigma_\tau$. Correlated errors and limb darkening cause
these formulas to be underestimates.

\bigskip
\bigskip
\centerline{\textbf{ 5. SUMMARY OF RECENT ACHIEVEMENTS}}
\bigskip

\noindent
\textbf{ 5.1 Discoveries of transiting planets}
\bigskip

\begin{figure*}
  \epsscale{2}
  \plotone{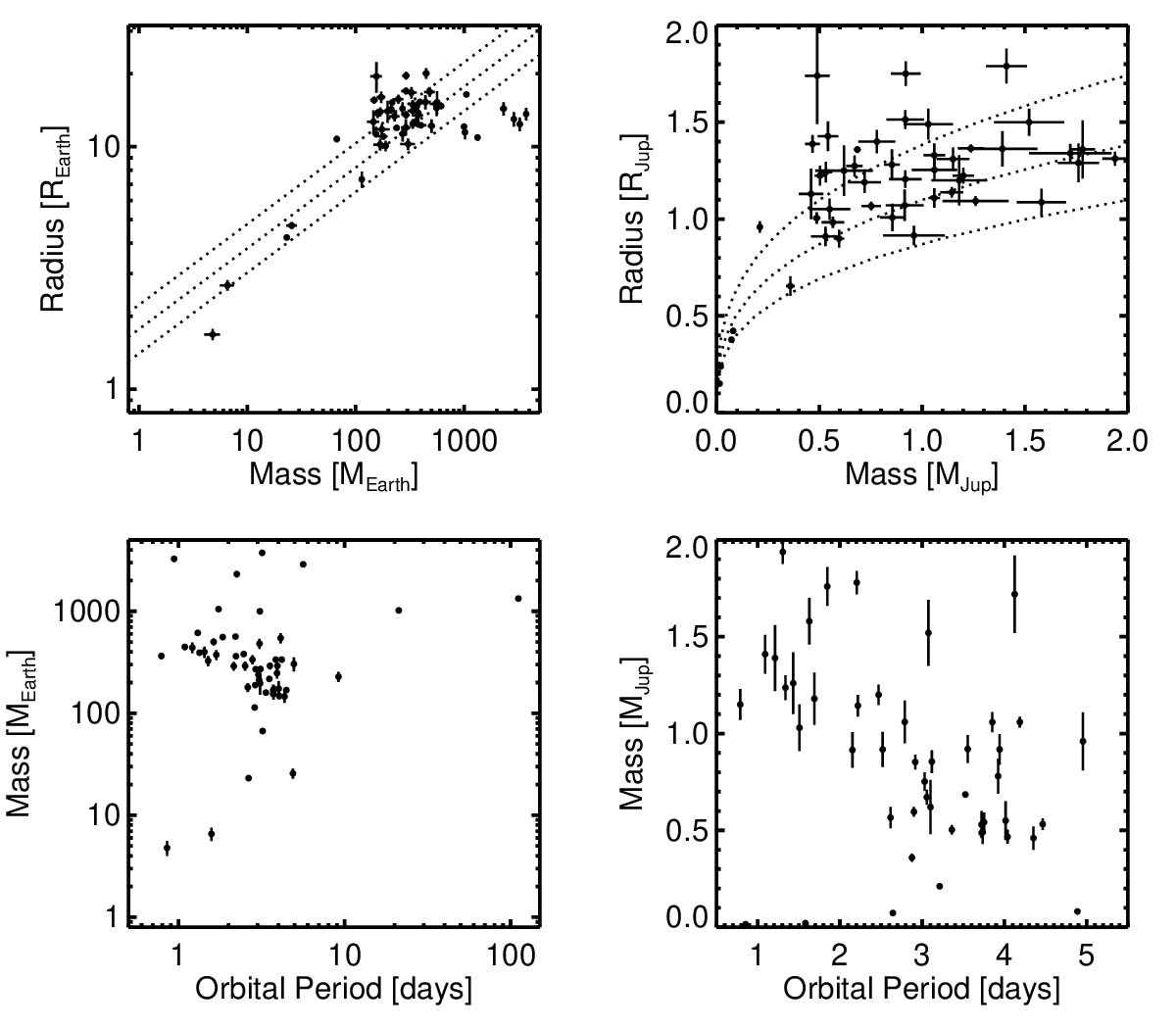}
  \caption{\small Masses, radii, and orbital periods of the transiting
    exoplanets. {\it Upper left:} Radius versus mass, on logarithmic
    scales. The dotted lines are loci of constant mean density (0.5,
    1, and 2~g~cm$^{-3}$, from top to bottom).  {\it Upper right:}
    Same, but on linear scales, and with the axes restricted to
    highlight the gas giants that dominate the sample. {\it Lower
      left:} Mass versus orbital period, on a logarithmic scale. The
    two long-period outliers are HD~17156b ($P=21$~d) and HD~80606b
    ($P=111$~d). {\it Lower right:} Same, but on a linear scale, and
    with axes restricted to highlight the gas giants. The
    anticorrelation between mass and orbital period is
    evident.\label{fig:mrp}}
\end{figure*}

Discoveries of exoplanets, and transiting exoplanets in particular,
have abounded in recent years (Figure~\ref{fig:disc-hist}). As of
December 2009, approximately 64 transiting planets are known,
representing 15\% of the total number of exoplanets
discovered. Figure~\ref{fig:mrp} shows their masses, radii, and
orbital periods. It is important to remember that these planets are
{\it not} a randomly selected subset of exoplanets. The properties of
the ensemble have been shaped by powerful selection effects in the
surveys that led to their discovery, favoring large planets in
short-period orbits.

Despite these selection effects, the known transiting planets exhibit
a striking diversity. They span three orders of magnitude in mass, and
one order of magnitude in radius. Most are gas giants, comparable in
mass and radius to Jupiter. There are also two planets with sizes more
like Neptune and Uranus, as well as two even smaller planets with
sizes only a few times larger than Earth, a category that has come to
be known as ``super-Earths.''

\begin{figure*}
  \epsscale{1.75}
  \plotone{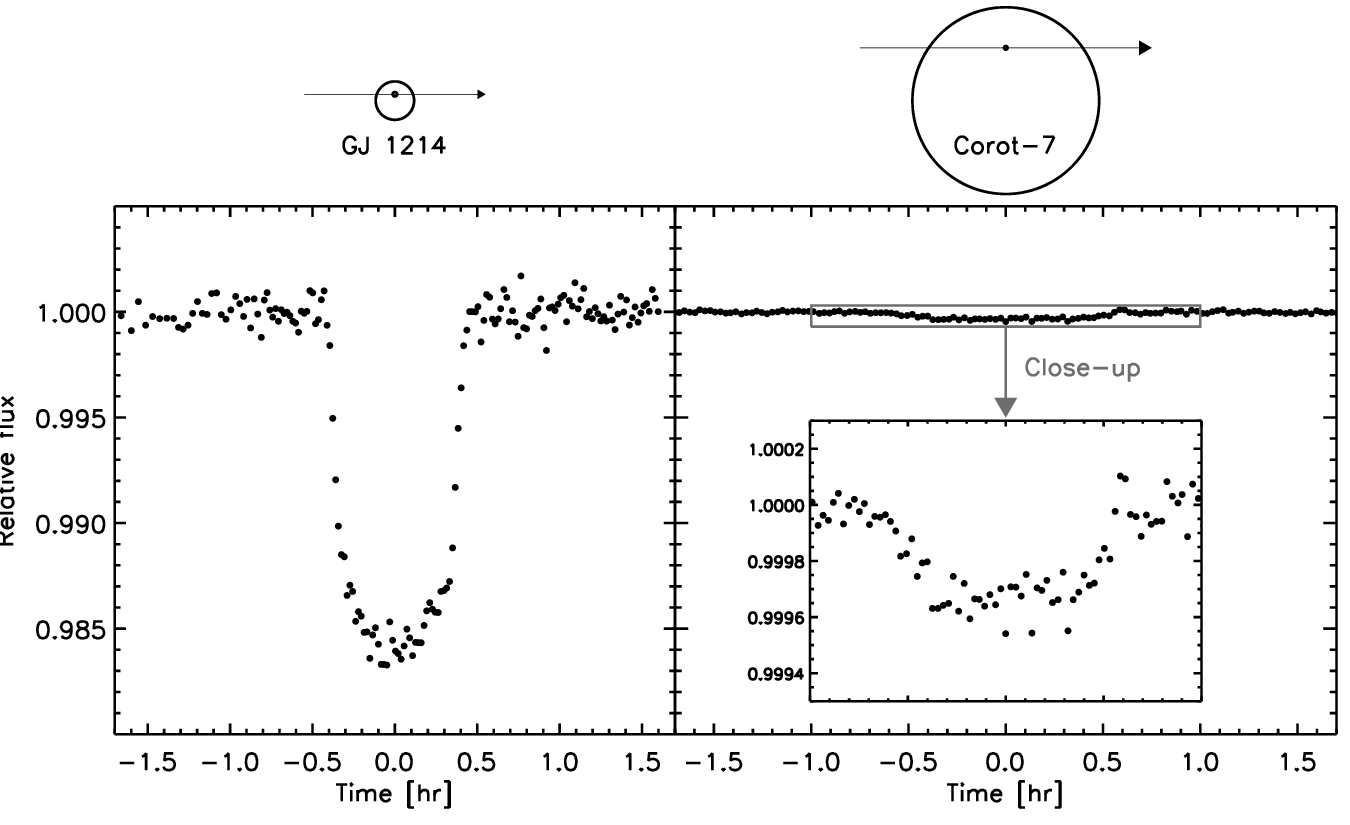}
  \caption{\small Transits of two different super-Earths, GJ~1214b
    (left) and CoRoT-7b (right). The planets are approximately the
    same size, but because GJ~1214b orbits a small star (spectral type
    M4.5V) its transit depth is much larger than that of CoRoT-7b,
    which orbits a larger star (G9V). References: Charbonneau et
    al.~(2009), L{\'e}ger et al.~(2009). \label{fig:smallplanets}}
\end{figure*}

The two transiting super-Earths (Figure~\ref{fig:smallplanets}) were
found with completely different strategies. The {\it CoRoT}
(COnvection, ROtation, and planetary Transits) team uses a satellite
equipped with a 0.27~m telescope and CCD cameras to examine fields of
$\sim$10,000 stars for a few months at a time, seeking relatively
short-period planets ($P\lesssim20$~d). Along with several giant
planets they have found a planet of radius 1.7~$R_\oplus$ in a 20-hour
orbit around a G dwarf star, producing a transit depth of only
$3.4\times 10^{-4}$ (L{\'e}ger et al.~2009). This is smaller than the
detection threshold of any of the ground-based surveys, demonstrating
the advantage of space-based photometry. However, it proved difficult
to spectroscopically confirm that the signal is indeed due to a
planet, because the host star is relatively faint and
chromospherically active (Queloz et al.~2009).

Another project, called MEarth (and pronounced {\it mirth}), seeks
transits of small planets from the ground by focusing on very small
stars (M dwarfs), for which even a super-Earth would produce a transit
depth of order 1\%. Because such stars are intrinsically faint, one
must search the whole sky to find examples bright enough for follow-up
work. That is why MEarth abandoned the usual survey concept in which
many stars are monitored within a single telescope's field of view.
Instead they monitor M dwarf stars one at a time, using several 0.4~m
telescopes. Using this strategy they found a planet of radius
$2.7~R_\oplus$ in a 1.6-day period around a star of radius
$0.21~R_\odot$ (Charbonneau et al.~2009). The large transit depth of
1.3\% invites follow-up observations to study the planet's atmosphere.

At the other extreme, several transiting objects have masses greater
than 10~$M_{\rm Jup}$, reviving the old debate about what should and
should not be considered a planet. The radii of these massive objects
are not much larger than Jupiter's radius, in agreement with
predictions that between about 1--50~$M_{\rm Jup}$ the pressure due to
Coulomb forces (which would give $R \propto M^{1/3}$) and electron
degeneracy pressure ($R \propto M^{-1/3}$) conspire to mute the
mass-dependence of the radius.

Almost all of the transiting planets have short orbital periods
($<$10~days), due to the decline in transit probability with orbital
distance (equation~\ref{eq:transit-probability-ecc}). Two conspicuous
exceptions are HD~17156b with $P=21$~days (Barbieri et al.~2009), and
HD~80606b with $P=111$~days (Moutou et al.~2009, Garcia-Melendo \&
McCullough~2009, Fossey et al.~2009). Both of those planets were
discovered by the Doppler method and found to transit through
photometric follow-up observations. They were recognized as
high-priority targets because both systems have highly eccentric
orbits oriented in such a way as to enhance the probability of
eclipses. Thus despite their long orbital periods, their periastron
distances are small ($<$0.1~AU) along with all of the other known
transiting planets.

\vspace{1in}
\bigskip
\noindent
\textbf{ 5.2 Follow-up photometry and absolute dimensions}
\bigskip

Follow-up observations of transits have allowed the basic transit
parameters $\{\delta, T, \tau\}$ to be determined to within 1\% or
better, and absolute dimensions of planets to within about 5\%.
Compilations of system parameters are given by Torres, Winn, \& Holman
(2008) and Southworth~(2009). The orbital periods are known with 8
significant digits in some cases. Ground-based observations have
achieved a photometric precision of 250~ppm per 1~min sample, through
the techniques described in Section 4.2.

\begin{figure*}
  \epsscale{1.21}
  \plotone{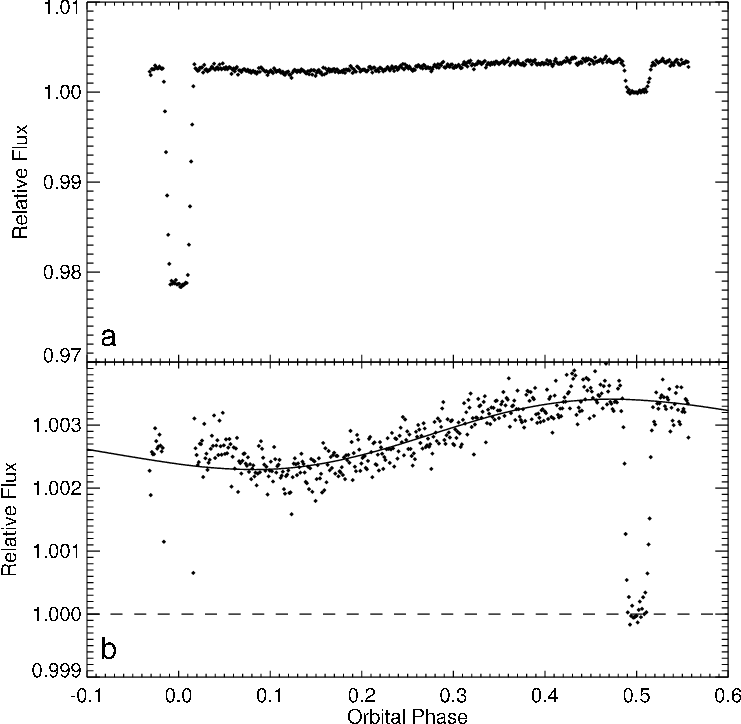}
  \caption{\small The combined 8~$\mu$m brightness of the K star
    HD~189733 and its giant planet, over a 33~hr interval including a
    transit and an occultation. The bottom panel shows the same data
    as the top panel but with a restricted vertical scale to highlight
    the gradual rise in brightness as the planet's dayside comes into
    view. The amplitude of this variation gives the temperature
    contrast between the dayside (estimated as $1211\pm 11$~K) and the
    nightside ($973\pm 33$~K). From Knutson et al.~(2007b).
    \label{fig:knutson-phasevar}}
\end{figure*}

A pathbreaking achievement was the {\it Hubble Space Telescope} light
curve of HD~209458 by Brown et al.~(2001), with a time sampling of
80~s and a precision of 110~ppm, without the need for comparison
stars. With the {\it Spitzer Space Telescope}\, the photon noise is
generally higher, but there are compensatory advantages: there is
little limb-darkening at at mid-infrared wavelengths, and
uninterrupted views of entire events are possible because the
satellite is not in a low-earth orbit. Among the most spectacular data
yet obtained in this field was the 33~hr observation by Knutson et
al.~(2007b) of the K star HD~189733, spanning both a transit and
occultation of its giant planet. The data were gathered with {\it
  Spitzer} at a wavelength of 8~$\mu$m, and are shown in
Figure~\ref{fig:knutson-phasevar}. They not only provided extremely
precise light curves of the transit and occultation, but also showed
that the combined flux rises gradually in between those two events,
demonstrating that the dayside is hotter than the nightside.

Among the properties of the close-in giant planets, a few patterns
have emerged. The planetary mass is inversely related to the orbital
period (Mazeh, Zucker, \& Pont 2005). This anticorrelation has been
variously attributed to selection effects, tidal interactions, and
thermal evaporation of the planetary atmosphere. There is also a
positive correlation between the metallicity of the host star and the
inferred ``core mass'' of the planet (Guillot et al.~2006), by which
is meant the mass of heavy elements required in models of the
planetary interior that agree with the observed mass and radius. It is
tempting to interpret this latter correlation as support for the
core-accretion theory of planet formation.

A persistent theme in this field, and a source of controversy and
speculation, is that several of the transiting giant planets have
radii that are 10-50\% larger than expected from models of
hydrogen-helium planets, even after accounting for the intense stellar
heating and selection effects (see the chapter by Correia and
Laskar). Among the possible explanations for these ``bloated'' planets
are tidal heating (see, e.g., Miller et al.~2009); unknown atmospheric
constituents that efficiently trap internal heat (Burrows et
al.~2007); enhanced downward convection allowing the incident stellar
radiation to reach significant depth (Guillot \& Showman 2002); and
inhibition of convection within the planet that traps internal heat
(Chabrier \& Baraffe 2007).

Likewise, a few planets are observed to be {\it smaller} than expected
for a hydrogen-helium planet with the observed mass and degree of
irradiation. Some examples are HD~149026b (Sato et al.~2005) and
HAT-P-3b (Torres et al.~2007). The favored interpretation is that
these planets are enriched in elements heavier than hydrogen and
helium, and the increased mean molecular weight leads to a larger
overall density. Some degree of enrichment might be expected because
Jupiter and Saturn are themselves enriched in hevay elements relative
to the Sun. However the two planets just mentioned would need to be
enriched still further; in the case of HD~149026b, theoretical models
suggest that the planet is 65\% heavy elements by mass.

\bigskip
\noindent
\textbf{ 5.3 Atmospheric physics}
\bigskip

Atmospheric spectra have been obtained for several gas giant planets,
especially the two ``hot Jupiters'' with the brightest host stars,
HD~209458b and HD~189733b. Both transmission and emission spectroscopy
have been undertaken, mainly with the space telescopes {\it Hubble}
and {\it Spitzer}. Figures~12 and 13 show examples of both
transmission (transit) and emission (occultation) spectra. These
observations have stimulated much theoretical work on exoplanetary
atmospheres (see the chapters by Chambers and by D'Angelo et al.\ for
more details than are presented here).

This enterprise began with the optical detection of neutral sodium in
a transmission spectrum of HD~209458b by Charbonneau et al.~(2002).
For that same planet, Vidal-Madjar et al.~(2003) measured an enormous
transit depth of $(15\pm 4)$\% within an ultraviolet bandpass
bracketing the wavelength of the Lyman-$\alpha$ transition, which they
attributed to neutral hydrogen gas being blown off the planet. For
HD~189733b, a rise in transit depth toward shorter wavelengths was
observed by Pont et al.~(2007) and interpreted as Rayleigh scattering
in the planet's upper atmosphere.

Recent attention has turned to infrared wavelengths, where molecules
make their imprint. In the emission spectra, departures from a
blackbody spectrum have been interpreted as arising from water,
methane, carbon monoxide, and carbon dioxide (see, e.g., Grillmair et
al.~2008, Swain et al.~2009). Absorption features in transmission
spectra have been interpreted as arising from water and methane (see,
e.g., Swain et al.~2008, Tinetti et al.~2007).

\begin{figure*}
  \centerline{\includegraphics[scale=0.55,angle=90]{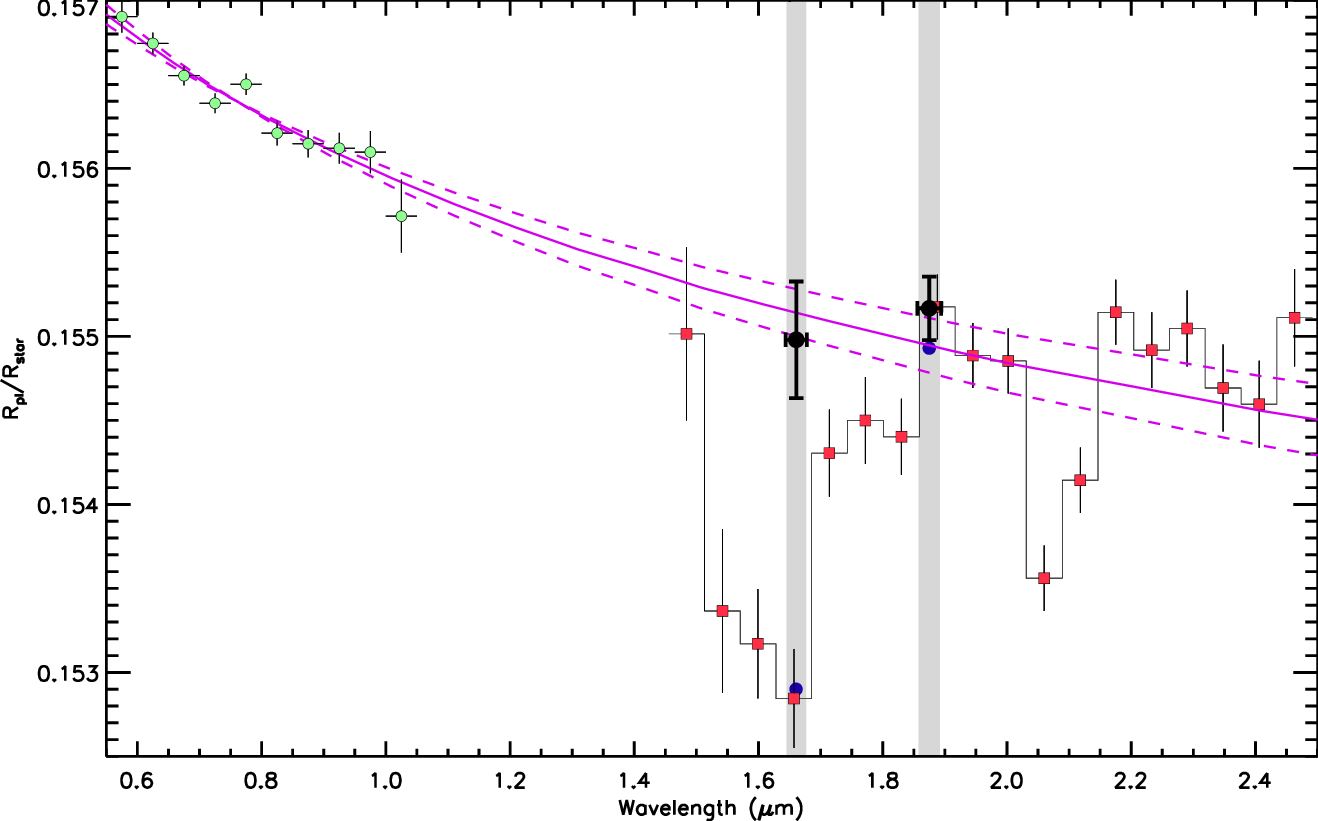}}
  \caption{\small Transmission (transit) spectroscopy of the gas giant
    HD~189733b, using the {\it Hubble Space Telescope}. The symbols
    with errors bars are measurements of the effective planet-to-star
    radius ratio as a function of wavelength. The dip at 1.6~$\mu$m
    was interpreted as evidence for water, and the rise at 2.1~$\mu$m
    as evidence for methane (Swain et al.~2008). However, subsequent
    observations at 1.7~$\mu$m and 1.9~$\mu$m, shown with darker
    symbols and gray bands, disagree with the earlier results and are
    consistent with a Rayleigh scattering model (solid and dashed
    curves). From Sing et al.~(2009).\label{fig:sing}}
\end{figure*}

\begin{figure*}
  \epsscale{1.35}
  \plotone{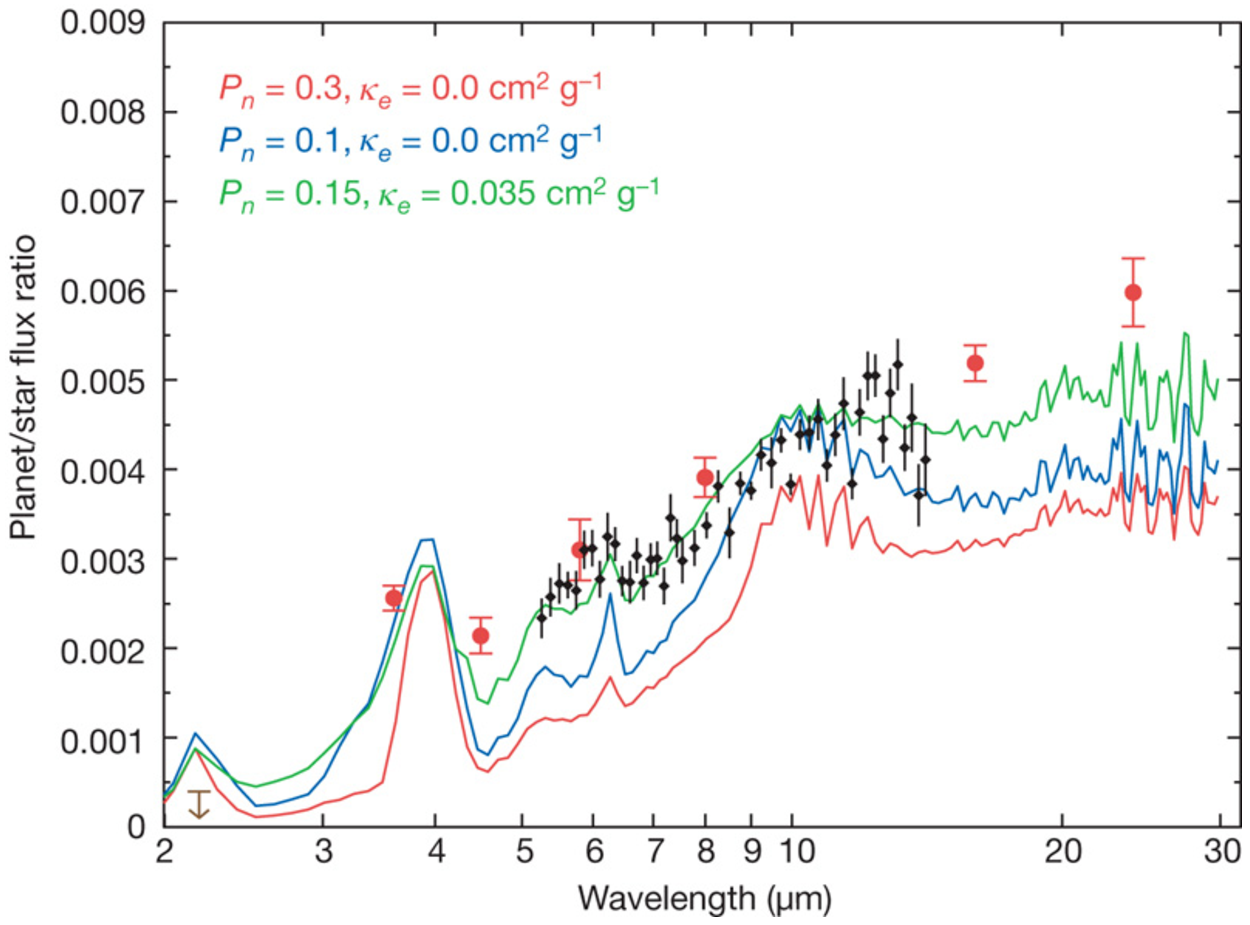}
  \caption{\small Occultation spectroscopy of the gas giant
    HD~189733b, using the {\it Spitzer Space Telescope}.  The points
    show measurements of the flux density ratio of the planet and star
    as a function of wavelength.  The smaller and finer-sampled points
    are based on observations with a dispersive spectrograph while the
    larger points are based on broadband filter photometry.  The three
    lines show the outputs of model atmospheres with varying choices
    for the parameters $P_n$, specifying the efficiency of heat
    transfer from dayside to nightside, and $\kappa_e$, specifying the
    opacity of a putative high-altitude absorbing species. From
    Grillmair et al.~(2008).\label{fig:grillmair}}
\end{figure*}

Despite these impressive achievements, many issues regarding the
atmospheres of hot Jupiters remain unsettled. Controversies arise
because spectral features are observed with a low signal-to-noise
ratio and are not always reproducible, as illustrated in
Figure~\ref{fig:sing}. This is understandable, as the observers have
pushed the instruments beyond their design specifications to make
these demanding measurements. In addition, the theoretical
interpretation of the spectra is in a primitive state. A published
spectrum is usually accompanied by a model that fits the data, but
left unanswered are whether the model is unique and how
well-constrained are its parameters. Recent work by Madhusudhan \&
Seager (2009) addresses this problem.

One theme that has arisen in the last few years is that some hot
Jupiters have an {\it inversion layer} in their upper atmospheres,
within which the temperature rises with height instead of the usual
decline. The evidence for an inversion layer is emission in excess of
a blackbody spectrum between 4--8~$\mu$m, where excess {\it
  absorption} was expected due to water vapor. The interpretation is
that water is seen in emission because it exists in a hot, tenuous
stratosphere.

Hot stratospheres develop when starlight is strongly absorbed by some
species at low pressure (at high altitude) where the atmosphere does
not radiate efficiently. The identity of this absorber in hot Jupiter
atmospheres has been a topic of debate. Gaseous titanium oxide and
vanadium oxide are candidates (Hubeny et al.~2003), as are
photochemically-produced sulfur compounds (Zahnle et
al.~2009). Meanwhile observers are searching for correlations between
the presence or absence of an inversion layer, the degree of stellar
irradiation, the magnitude of the day-night temperature difference,
and other observable properties.

A few of the known planets have highly eccentric orbits and small
periastron distances, and are therefore subject to highly variable
stellar irradiation. Observers have monitored the planetary thermal
emission following periastron passages, to gauge the amplitude and
timescale of the thermal response to stellar heating. In the most
extreme case, Laughlin et al.~(2009) watched the effective temperature
of the giant planet HD~80606b rise from about 800~K to 1500~K after
periastron passage, and inferred that the characteristic radiative
timescale of the upper atmosphere is about 4.5~hr.

Although the thermal emission from hot Jupiters has been detected in
many cases, at infrared and optical wavelengths, there has been no
unambiguous detection of starlight {\it reflected} from the planetary
atmosphere. The best resulting upper limit on the visual planetary
albedo is 0.17 (with 3$\sigma$ confidence), for the case of HD~209458b
(Rowe et al.~2008). This rules out highly reflective clouds of the
sort that give Jupiter its visual albedo of 0.5. However, the limits
have not been of great interest to the theorists, who predicted all
along that the visual albedos of hot Jupiters would be very small.

\bigskip
\noindent
\textbf{ 5.4 Tidal evolution and migration}
\bigskip

Almost all of the known transiting planets are close enough to their
parent stars for tidal effects to be important. The tidal bulges on
the star and planet provide ``handles'' for the bodies to torque each
other, and tidal friction slowly drains energy from the system (see
the chapter by Fabrycky). For a typical hot Jupiter, the sequence of
events is expected to be as follows: (i) Over $\sim$10$^6$~yr, the
planet's rotational period is synchronized with its orbital period,
and its obliquity (the angle between its rotational and orbital
angular momentum vectors) is driven to zero. (ii) Over
$\sim$10$^9$~yr, the orbit is circularized. (iii) After
$\sim$10$^{12}$~yr (i.e.\ not yet), the stellar rotational period is
synchronized with the orbital period, the stellar obliquity is driven
to zero, and the orbit decays, leading to the engulfment of the
planet. The timescales for these processes are highly uncertain, and
more complex histories are possible if there are additional planets in
the system or if the planet's internal structure is strongly affected
by tidal heating.

Tidal circularization is implicated by the fact that planets with
orbital periods shorter than $\sim$10~days tend to have smaller
orbital eccentricities than longer-period planets. This fact was
already known from Doppler surveys, but for eclipsing planets the
eccentricities can be measured more precisely (see Section 3.2). It is
possible that the ``bloating'' of some of the close-in giant planets
(Section 5.3) is related to the heat that was produced during the
circularization process, or that may still be ongoing.

As for tidal synchronization, observations with {\it Spitzer} have
revealed planets with a cold side facing away from the star and a hot
side facing the star (see Figure~\ref{fig:knutson-phasevar}). This
could be interpreted as evidence for synchronization, although it is
not definitive, because it is also possible that the heat from the
star is re-radiated too quickly for advection to homogenize the upper
atmosphere of the planet.

Tidal decay of the orbit would be observable as a gradual decline in
the orbital period (Sasselov 2003). For most systems the theoretical
timescale for tidal decay is much longer than the age of the star, and
indeed no evidence for this process has been found. However in at
least one case (WASP-18b; Hellier et al.~2009) the theoretical
timescale for tidal decay is much {\it shorter}\, than the stellar
age, because of the large planetary mass and short orbital period.
The existence of this system suggests that the theoretical
expectations were wrong and dissipation is slower in reality.

Likewise, in most cases one would not expect that enough time has
elapsed for tides to modify the star's spin rate or orientation
(Barker \& Ogilvie 2009). This suggests that the measurements of the
projected spin-orbit angle using the RM effect (Section 3.5) should be
interpreted in the context of planet formation and evolution rather
than tides. A close spin-orbit alignment is expected because a star
and its planets inherit their angular momentum from a common source:
the protostellar disk. However, for hot Jupiters there is the
complication that they presumably formed at larger distances and
``migrated'' inward through processes that are poorly understood. Some
of the migration theories predict that the original spin-orbit
alignment should be preserved, while others predict occasionally large
misalignments. For example, tidal interactions with the protoplanetary
disk (see chapter by Showman et al.) should drive the system into
close alignment (Marzari \& Nelson 2009), while planet-planet
scattering or Kozai oscillations with tidal friction (see chapter by
Fabrycky) should result in misaligned systems.

\begin{figure*}
  \epsscale{2.0}
  \plotone{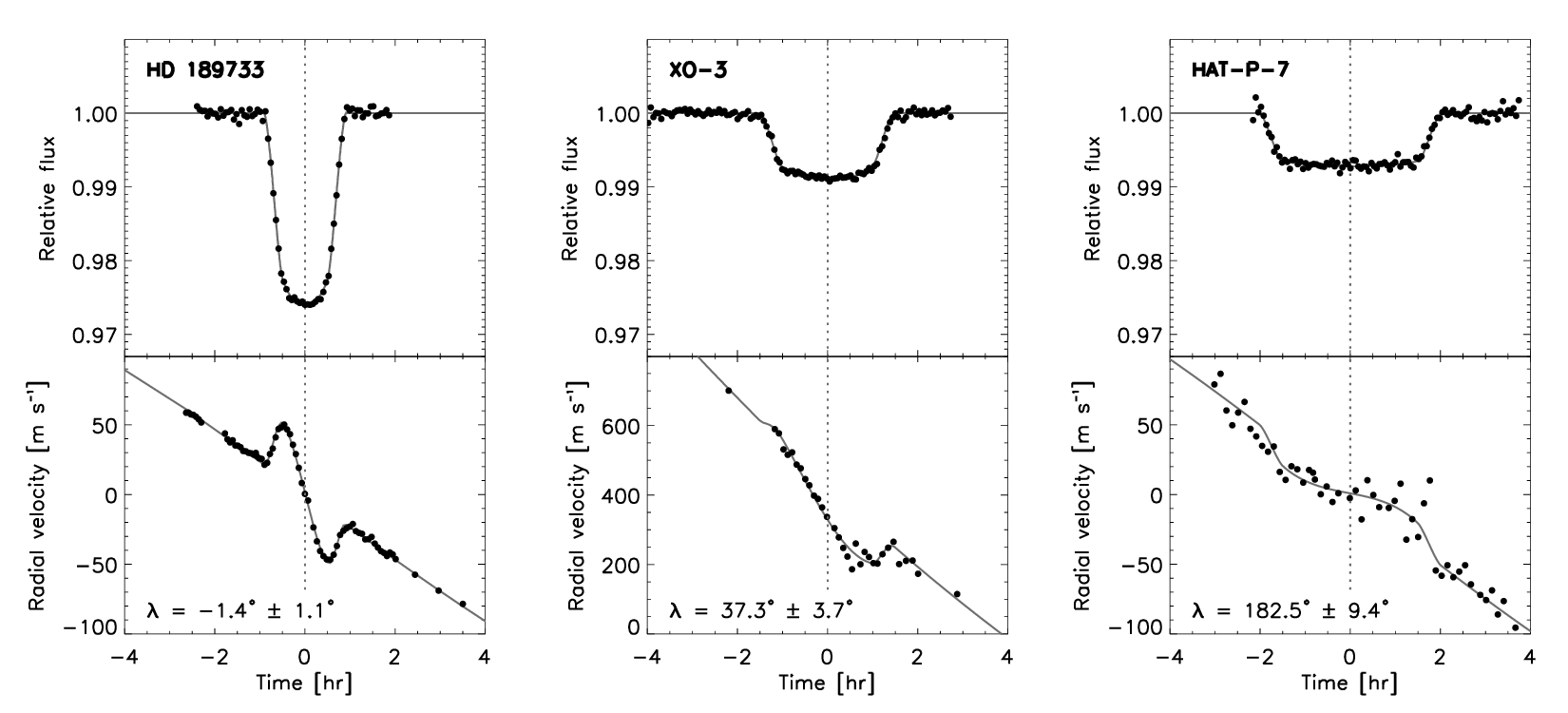}
  \caption{\small Examples of data used to measure the projected
    spin-orbit angle $\lambda$. The top panels show transit
    photometry, and the bottom panels show the apparent radial
    velocity of the star, including both orbital motion and the
    anomalous Doppler shift (the Rossiter-McLaughlin effect). The left
    panels show a well-aligned system and the middle panels show a
    misaligned system. The right panels show a system for which the
    stellar and orbital ``north poles'' are nearly {\it antiparallel}
    on the sky, indicating that the planet's orbit is either
    retrograde or polar (depending on the unknown inclination of the
    stellar rotation axis). References: Winn et al.~(2006;
    2009a,b).\label{fig:rossiter_data}}
\end{figure*}

The projected spin-orbit angle has been measured for about 20
exoplanets, all of them close-in giants. Some examples of data are
shown in Figure~\ref{fig:rossiter_data}. In many cases the results are
consistent with good alignment, with measurement precisions ranging
from 1--20$^\circ$ (for a recent summary, see Fabrycky \& Winn
2009). However there are now at least 4 clear cases of misaligned
systems. One such case is XO-3b, a massive planet in a close-in
eccentric orbit that is tilted by more than 30$^\circ$ with respect to
the stellar equator (H{\'e}brard et al.~2008, Winn et al.~2009a). Even
more dramatic is HAT-P-7b, for which the planetary orbit and stellar
spin axis are tilted by more than $86^\circ$ (Winn et al.~2009b,
Narita et al.~2009). The planetary orbit is either polar (going over
the north and south poles of the star) or retrograde (revolving in the
opposite direction as the star is rotating). Another possible
retrograde system is WASP-17b (Anderson et al.~2009).

\bigskip
\centerline{\textbf{ 6. FUTURE PROSPECTS}}
\bigskip

Eclipses are the ``here and now'' of exoplanetary science. It seems
that every month, someone reports a startling observational feat,
describes a creative new application of eclipse data, or proposes an
ambitious survey to find ever-larger numbers of ever-smaller
transiting planets. Keeping up with the field leaves one breathless,
and wary of making predictions.

The only safe bet is that the {\it Kepler} space mission will have a
major impact. In March 2009, the {\it Kepler} satellite was launched
into an an Earth-trailing orbit where it will stare at a single field
of $10^5$ stars for at least 3.5~yr, seeking Earth-like planets in
Earth-like orbits around Sun-like stars (Borucki et al.~2003). {\it
  Kepler} will end our obsession with close-in, tidally-influenced,
strongly-irradiated giant planets. At last it will be possible to
study eclipses of multitudes of longer-period and smaller-radius
planets. Stellar heating will not confound models of their atmospheres
or interiors, and tidal effects will not confound the interpretation
of their orbital properties.

Many of {\it Kepler}'s discoveries will be small enough to have solid
surfaces, making it possible to measure masses and radii of rocky
exoplanets. The mission's highest priority goal is to determine the
abundance of terrestrial planets in the ``habitable zones'' of their
parent stars, defined as the range of orbital distances within which
water could exist in liquid form on the surface of a rocky planet. The
abundance of such planets is a key input to the ``Drake equation'' as
well as to theories of planet formation and to the designs of future
missions that will find and study the nearest examples.

The precision of space-based photometry will also be put to good use
to detect a host of subtle effects that have been described in the
literature but not yet observed. Already, data from the {\it Kepler}
and {\it CoRoT} satellites have been used to detect occultations in
visible light, as opposed to infrared (Borucki et al.~2009, Snellen et
al.~2009), although in both cases the signal is attributed mainly to
thermal emission rather than the elusive reflected component.

Many investigators will seek to detect short-term variations in the
times, durations, and depths of transits due to gravitational
perturbations by additional planets, as mentioned in Section 3.2 and
discussed further in the chapter by Fabrycky. In particular, {\it
  Kepler} might find habitable planets not only by detecting their
transits, but also by observing their effects on other transiting
planets. The same principle can be used to detect habitable
``exomoons'' (Kipping et al.~2009) or habitable planets at the Trojan
points (Lagrange L4 and L5) in the orbits of more massive planets
(Ford \& Gaudi 2006, Madhusudhan \& Winn 2009).

Longer-term orbital perturbations should also be measurable, due to
additional bodies (Miralda-Escud\'e 2002) or relativistic precession
(Jord{\'a}n \& Bakos 2008, P{\'a}l \& Kocsis~2008). For very close-in
planets, the orbital precession rate should be dominated by the
effects of the planetary tidal bulge, which in turn depends on the
deformability of the planet. This raises the prospect of using the
measured precession rate to infer some aspects of the planet's
interior structure (Ragozzine \& Wolf 2009).

In addition, the precise form of the transit light curve depends on
the shape of the planet's silhouette. With precise enough photometry
it may be possible to detect the departures from sphericity due to
rings (Barnes \& Fortney 2004) or rotation (Seager \& Hui 2002, Barnes
\& Fortney 2003). Already it has been shown that HD~189733b is less
oblate than Saturn (Carter \& Winn 2009b), as expected if the planet's
rotation period is synchronized with its 2.2-day orbit (slower than
Saturn's 11~hr rotation period). Soon we will have a sample of
transiting planets with larger orbital distances, for which
synchronization is not expected and which may therefore be rotating
quickly enough for the oblateness to be detectable.

Another prize that remains to be won is the discovery of a system with
more than one transiting planet. This would give empirical constraints
on the mutual inclination of exoplanetary orbits (Nesvorn{\'y} 2009),
as well as estimates of the planetary masses that are independent of
the stellar mass, through the observable effects of planet-planet
gravitational interactions. Already there are a few cases in which
there is evidence for a second planet around a star that is known to
have a transiting planet (see, e.g., Bakos et al.~2009), but in none
of those cases is it known whether the second planet also transits.

Ground-based transit surveys will still play an important role by
targeting brighter stars over wider fields than {\it CoRoT} and {\it
  Kepler}. With brighter stars it is easier to rule out false
positives, and to measure the planetary mass through radial-velocity
variations of the host star. For the smallest planets around the
relatively faint stars in the {\it CoRoT} and {\it Kepler} fields this
will be difficult. Bright stars also offer more photons for the
high-precision follow-up investigations that make eclipses so
valuable.

Ground-based surveys will continue providing such targets, and in
particular will mine the southern sky, which is comparatively
unexplored. Two contenders are the SuperWASP and HAT-South surveys,
which will aim to improve their photometric precision and discover
many Neptune-sized or even smaller planets in addition to the gas
giants. The MEarth project, mentioned in Section 5.1, is specifically
targeting small and low-luminosity stars because of the less stringent
requirements on photometric precision and because the habitable zones
occur at smaller orbital distances, making transits more likely and
more frequent (Nutzman \& Charbonneau 2008).

An attractive idea is to design a survey with the fine photometric
precision that is possible from space, but that would somehow survey
the brightest stars on the sky instead of those within a narrow field
of view. Several mission concepts are being studied: the {\it
  Transiting Exoplanet Survey Satellite} (Deming et al.~2009) and the
{\it PLATO} mission (Lindberg et al.~2009) would tile the sky with the
fields-of-view of small cameras, while {\it LEAVITT} would have a
series of telescopes on a spinning platform (Olling 2007). Another
concept is to deploy an armada of ``nanosatellites'' to target
individual stars (S.~Seager, priv.\ comm.\, 2009).

Each time a terrestrial planet is found transiting a bright star, a
period of intense anticipation will ensue, as astronomers try to
characterize its atmosphere through transmission or emission
spectroscopy. The anticipation will be especially keen for any planets
in the habitable zone of their parent stars, which may reveal water or
other molecules considered important for life. Attaining the necessary
signal-to-noise ratio will be excruciating work, and may require the
{\it James Webb Space Telescope} or an even more capable
special-purpose instrument.

Transits and occultations have had a disproportionate impact early in
the history of exoplanetary science. This is mainly because of the
unexpected existence of close-in giant planets, and their high transit
probabilities. In the coming years, other techniques such as
astrometry, microlensing, and spatially-resolved imaging will mature
and give a more complete accounting of exoplanetary systems. Still, in
closing, let us recall that most of our most fundamental and precise
information about {\it stars} comes from eclipsing systems, even after
more than a century of technological development since eclipses were
first observed. The same is likely to be true for exoplanets.

\bigskip

\textbf{ Acknowledgments.} The author is grateful to Simon Albrecht,
Josh Carter, Dan Fabrycky, David Kipping, Heather Knutson, Jack
Lissauer, Geoff Marcy, Frederic Pont, Sara Seager, Eric Agol, George
Ricker, Geza Kovacs, and the anonymous referees, for helpful comments
on this chapter.

\bigskip

\centerline{\textbf{ REFERENCES}}
\bigskip
\parindent=-10pt
\parskip=0pt
{\small
\baselineskip=11pt

\input references.tex

\end{document}

%% file: references.tex
\refs Agol, E., Steffen, J., Sari, R., Clarkson, W.\ (2005) On
detecting terrestrial planets with timing of giant planet transits.\
{\em Mon.\ Not.\ Roy.\ Astr.\ Soc.\ 359}, 567-579.

\refs Alonso, R., and 11 colleagues (2004) TrES-1: The Transiting
Planet of a Bright K0 V Star.\ {\it Astrophys.\ J.\ 613}, L153-L156.

\refs Anderson, D.~R., and 20 colleagues (2010) WASP-17b: An Ultra-Low
Density Planet in a Probable Retrograde Orbit.\ {\it Astrophysical
Journal 709}, 159-167.
 
\refs Baines, E.~K., van Belle, G.~T., ten Brummelaar, T.~A.,
McAlister, H.~A., Swain, M., Turner, N.~H., Sturmann, L., Sturmann,
J.\ (2007) Direct Measurement of the Radius and Density of the
Transiting Exoplanet HD 189733b with the CHARA Array.\ {\it Astroph.\
J.\ 661}, L195-L198.
 
\refs Bakos, G.~{\'A}., and 18 colleagues (2009) HAT-P-13b,c: A
Transiting Hot Jupiter with a Massive Outer Companion on an Eccentric
Orbit.\ {\it Astrophysical Journal 707}, 446-456.

\refs Bakos, G.~{\'A}., and 18 colleagues (2007) HAT-P-1b: A
Large-Radius, Low-Density Exoplanet Transiting One Member of a Stellar
Binary.\ {\it Astrophys.\ J.\ 656}, 552-559.
 
\refs Barbieri, M., and 10 colleagues (2007) HD 17156b: a transiting
planet with a 21.2-day period and an eccentric orbit.\ {\it Astronomy
and Astrophysics 476}, L13-L16.

\refs Barker, A.~J., Ogilvie, G.~I.\ (2009) On the tidal evolution of
Hot Jupiters on inclined orbits.\ {\em Mon.\ Not.\ Roy.\ Astr.\ Soc.\
395}, 2268-2287.
 
\refs Barnes, J.~W., Fortney, J.~J.\ (2003) Measuring the Oblateness
and Rotation of Transiting Extrasolar Giant Planets.\ {\it
Astrophysical Journal 588}, 545-556.

\refs Barnes, J.~W., Fortney, J.~J.\ (2004) Transit Detectability of
Ring Systems around Extrasolar Giant Planets.\ {\it Astrophysical
Journal 616}, 1193-1203.

\refs Beatty, T.~G., \& Gaudi, B.~S.\ (2008) Predicting the Yields of
Photometric Surveys for Transiting Extrasolar Planets, {\em Astroph.\
J., 686}, 1302-1330.
 
\refs Borucki, W.~J., and 24 colleagues (2009) Kepler's Optical Phase
Curve of the Exoplanet HAT-P-7b.\ {\it Science 325}, 709.

\refs Brown, T.~M.\ (2001) Transmission Spectra as Diagnostics of
Extrasolar Giant Planet Atmospheres.\ {\it Astroph.\ J.\ 553},
1006-1026.

\refs Brown, T.~M., Charbonneau, D., Gilliland, R.~L., Noyes, R.~W.,
Burrows, A.\ (2001) Hubble Space Telescope Time-Series Photometry of
the Transiting Planet of HD 209458.\ {\it Astrophysical Journal 552},
699-709.

\refs Borucki, W.~J., and 12 colleagues (2003) The Kepler mission: a
wide-field-of-view photometer designed to determine the frequency of
Earth-size planets around solar-like stars.\ {\it Soc.\ Photo-Optical
Instr.\ Eng.\ (SPIE) Conf.\ Ser.\ 4854}, 129-140.

\refs Burke, C.~J., and 17 colleagues (2007) XO-2b: Transiting Hot
Jupiter in a Metal-rich Common Proper Motion Binary.\ {\it
Astrophysical Journal 671}, 2115-2128.
 
\refs Burrows, A., Hubeny, I., Budaj, J., Hubbard, W.~B.\ (2007)
Possible Solutions to the Radius Anomalies of Transiting Giant
Planets.\ {\it Astrophys.\ J.\ 661}, 502-514.
 
\refs Carciofi, A.~C.\ \& Magalh\~aes, A.~M.\ (2005) The Polarization
Signature of Extrasolar Planet Transiting Cool Dwarfs.\ {\it
Astrophs.\ J.\ 635}, 570-577.

\refs Carter, J.~A., Yee, J.~C., Eastman, J., Gaudi, B.~S., Winn,
J.~N.\ (2008) Analytic Approximations for Transit Light-Curve
Observables, Uncertainties, and Covariances.\ {\it Astrophysical
Journal 689}, 499-512.

\refs Carter, J.~A., Winn, J.~N., Gilliland, R., Holman, M.~J.\ (2009)
Near-Infrared Transit Photometry of the Exoplanet HD 149026b.\ {\it
Astrophysical Journal 696}, 241-253.

\refs Carter, J.~A., Winn, J.~N.\ (2009a) Parameter Estimation from
Time-series Data with Correlated Errors: A Wavelet-based Method and
its Application to Transit Light Curves.\ {\it Astrophysical Journal
704}, 51-67.

\refs Carter, J.~A., Winn, J.~N.\ (2009b) Empirical Constraints on the
Oblateness of an Exoplanet.\ {\it Astrophysical Journal}, 709, 1219.

\refs Chabrier, G., Baraffe, I.\ 2007.\ Heat Transport in Giant
(Exo)planets: A New Perspective.\ {\it Astrophysical Journal 661},
L81-L84.
 
\refs Charbonneau, D., Brown, T.~M., Latham, D.~W., Mayor, M.\ (2000)
Detection of Planetary Transits Across a Sun-like Star.\ {\it
Astrophys.\ J.\ 529}, L45-L48.
 
\refs Charbonneau, D., Brown, T.~M., Noyes, R.~W., Gilliland, R.~L.\
(2002) Detection of an Extrasolar Planet Atmosphere.\ {\it Astrophys.\
J.\ 568}, 377-384.
 
\refs Charbonneau, D.\ and 18 colleagues (2009) A super-Earth
transiting a nearby low-mass star. {\it Nature 462}, 891-894.
 
\refs Claret, A.\ (2004) A new non-linear limb-darkening law for LTE
stellar atmosphere models.\ III.\ Sloan filters: Calculations for
$-5.0 \leq $~log~[M/H]~$\le +1$, 2000~K~$\leq T_{\rm eff} \leq
50000$~K at several surface gravities.\ {\it Astron.\ Astroph.\ 428},
1001-1005.

\refs Claret, A.\ (2009) Does the HD 209458 planetary system pose a
challenge to the stellar atmosphere models? {\it Astronomy and
Astrophysics 506}, 1335-1340.

\refs Collier Cameron, A., and 31 colleagues (2007) Efficient
identification of exoplanetary transit candidates from SuperWASP light
curves.\ {\it Monthly Notices of the Royal Astronomical Society 380},
1230-1244.
 
\refs de Kort, J.~J.~M.~A.\ (1954) Upper and lower limits for the
eccentricity and longitude of periastron of an eclipsing binary.\ {\it
Ricerche Astronomiche 3}, 109-118.

\refs Deming, D., and 11 colleagues (2009) Discovery and
Characterization of Transiting Super Earths Using an All-Sky Transit
Survey and Follow-up by the James Webb Space Telescope.\ {\it
Publications of the Astronomical Society of the Pacific 121}, 952-967.
 
\refs Everett, M.~E.\ \& Howell, S.~B.\ (2001) A Technique for
Ultrahigh-Precision CCD Photometry.\ {\it Publ.\ Astron. Soc.\ Pacific
113}, 1428-1435.

\refs Fabrycky, D.~C., Winn, J.~N.\ (2009) Exoplanetary Spin-Orbit
Alignment: Results from the Ensemble of Rossiter-McLaughlin
Observations.\ {\it Astroph.\ J.\ 696}, 1230-1240.

\refs Ford, E.~B., Gaudi, B.~S.\ (2006) Observational Constraints on
Trojans of Transiting Extrasolar Planets.\ {\it Astrophysical Journal
652}, L137-L140.

\refs Fossey, S.~J., Waldmann, I.~P., Kipping, D.~M.\ (2009) Detection
of a transit by the planetary companion of HD 80606.\ {\it Monthly
Notices of the Royal Astronomical Society 396}, L16-L20.

\refs Garcia-Melendo, E., McCullough, P.~R.\ (2009) Photometric
Detection of a Transit of HD 80606b.\ {\it Astrophysical Journal 698},
558-561.

\refs Gaudi, B.~S.\ (2005) On the Size Distribution of Close-in
Extrasolar Giant Planets.\ {\it Astroph. J.\ 628}, L73-L76

\refs Gaudi, B.~S., Seager, S., \& Mallen-Ornelas, G.\ (2005) On the
Period Distribution of Close-in Extrasolar Giant Planets.\ {\it
Astroph.\ J.\ 623}, 472-481.
 
\refs Gaudi, B.~S., Winn, J.~N.\ (2007) Prospects for the
Characterization and Confirmation of Transiting Exoplanets via the
Rossiter-McLaughlin Effect.\ {\it Astroph.\ J.\ 655}, 550-563.
 
\refs Gilliland, R.~L., \& Brown, T.~M.\ (1988) Time-resolved CCD
photometry of an ensemble of stars.\ {\it Publ.\ Astron.\ Soc.\
Pacific 100}, 754-765.
 
\refs Gilliland, R.~L., and 23 colleagues (2000) A Lack of Planets in
47 Tucanae from a Hubble Space Telescope Search.\ {\it Astrophys.\ J.\
545}, L47-L51.

\refs Gillon, M., Demory, B.-O., Barman, T., Bonfils, X., Mazeh, T.,
Pont, F., Udry, S., Mayor, M., Queloz, D.\ (2007) Accurate Spitzer
infrared radius measurement for the hot Neptune GJ 436b.\ {\it
Astronomy and Astrophysics 471}, L51-L54.

\refs Gim{\'e}nez, A.\ (2007) Equations for the analysis of the light
curves of extra-solar planetary transits.\ {\it Astron.\ Astrophys.\
  474}, 1049-1049.

\refs Gregory, P.~C.\ (2005) {\it Bayesian Logical Data Analysis for
the Physical Sciences: A Comparative Approach with Mathematica
Support}, Cambridge University Press

\refs Grillmair, C.~J., Burrows, A., Charbonneau, D., Armus, L.,
Stauffer, J., Meadows, V., van Cleve, J., von Braun, K., Levine, D.\
(2008) Strong water absorption in the dayside emission spectrum of the
planet HD189733b.\ {\it Nature 456}, 767-769.
 
\refs Guillot, T., Santos, N.~C., Pont, F., Iro, N., Melo, C., Ribas,
I.\ (2006) A correlation between the heavy element content of
transiting extrasolar planets and the metallicity of their parent
stars.\ {\it Astron.\ Astroph.\ 453}, L21-L24.
 
\refs Guillot, T., Showman, A.~P.\ (2002) Evolution of ``51 Pegasus
b-like'' planets.\ {\it Astronomy and Astrophysics 385}, 156-165.
 
\refs H{\'e}brard, G., and 22 colleagues (2008) Misaligned spin-orbit
in the XO-3 planetary system? {\it Astron.\ Astrophys.\ 488}, 763-770.
 
\refs Hellier, C., and 22 colleagues (2009) An orbital period of 0.94
days for the hot-Jupiter planet WASP-18b.\ {\it Nature 460},
1098-1100.
 
\refs Henry, G.~W., Marcy, G.~W., Butler, R.~P., Vogt, S.~S.\ (2000) A
Transiting ``51 Peg-like'' Planet.\ {\it Astrophys.\ J.\ 529}, L41-L44

\refs Hilditch, R.~W.\ (2001) {\it An Introduction to Close Binary
Stars} (Cambridge, UK: Cambridge University Press)

\refs Holman, M.~J., Murray, N.~W.\ (2005) The Use of Transit Timing
to Detect Terrestrial-Mass Extrasolar Planets.\ {\it Science 307},
1288-1291.

\refs Holman, M.~J., Winn, J.~N., Latham, D.~W., O'Donovan, F.~T.,
Charbonneau, D., Bakos, G.~A., Esquerdo, G.~A., Hergenrother, C.,
Everett, M.~E., P{\'a}l, A.\ (2006) The Transit Light Curve
Project. I. Four Consecutive Transits of the Exoplanet XO-1b.\ {\it
Astrophysical Journal 652}, 1715-1723.

\refs Horne, K.\ (2003) Status aand Prospects of Planetary Transit
Searches: Hot Jupiters Galore.\ in {\it Scientific Frontiers in
Research on Extrasolar Planets}, ed.\ D.\ Deming and S.\ Seager, ASP
Conference Series Vol.\ 294 (San Francisco: ASP), p.\ 361-370.

\refs Howell, S.~B.\ (2006) {\it Handbook of CCD astronomy}
(Cambridge, UK: Cambridge Univ.\ Press), 2nd ed.
 
\refs Hubeny, I., Burrows, A., Sudarsky, D.\ (2003) A Possible
Bifurcation in Atmospheres of Strongly Irradiated Stars and Planets.\
{\it Astrophysical Journal 594}, 1011-1018.

\refs Jord{\'a}n, A., Bakos, G.~{\'A}.\ (2008)\ Observability of the
General Relativistic Precession of Periastra in Exoplanets.\ {\it
Astrophys.\ J.\ 685}, 543-552.

\refs Kallrath, J.\ \& Milone, E.~F.\ (2009) {\it Eclipsing Binary
Stars: Modeling and Analysis} (New York: Springer-New York), 2nd ed.,
in press

\refs Kane, S.~R.\ (2007) Detectability of exoplanetary transits from
radial velocity surveys. {\em Mon.\ Not.\ Roy.\ Astr.\ Soc.\ 380},
1488-1496.
 
\refs Kipping, D.~M.\ (2008) Transiting planets -- light-curve
analysis for eccentric orbits.\ {\em Mon.\ Not.\ Roy.\ Astr.\ Soc.\
389}, 1383-1390.
 
\refs Kipping, D.~M.\ (2009) Transit timing effects due to an
exomoon.\ {\em Mon.\ Not.\ Roy.\ Astr.\ Soc.\ 392}, 181-189.

\refs Kjeldsen, H.\ \& Frandsen, S.\ (1992) High-precision
time-resolved CCD photometry.\ {\it Publ.\ Astron.\ Soc.\ Pacific
104}, 413-434.
 
\refs Knutson, H.~A., Charbonneau, D., Noyes, R.~W., Brown, T.~M.,
Gilliland, R.~L.\ (2007a)\ Using Stellar Limb-Darkening to Refine the
Properties of HD 209458b.\ {\it Astrophys.\ J.\ 655}, 564-575.
 
\refs Knutson, H.~A., Charbonneau, D., Allen, L.~E., Fortney, J.~J.,
Agol, E., Cowan, N.~B., Showman, A.~P., Cooper, C.~S., Megeath, S.~T.\
(2007b) A map of the day-night contrast of the extrasolar planet HD
189733b.\ {\it Nature 447}, 183-186.
 
\refs Kopal, Z.\ (1979) Language of the stars: A discourse on the
theory of the light changes of eclipsing variables.\ {\it Astrophysics
and Space Science Library 77}, Kluwer, Dordrecht.

\refs Kov{\'a}cs, G., Zucker, S., Mazeh, T.\ (2002) A box-fitting
algorithm in the search for periodic transits.\ {\it Astronomy and
Astrophysics 391}, 369-377.
 
\refs Kov{\'a}cs, G., Bakos, G., Noyes, R.~W.\ 2005.\ A trend
filtering algorithm for wide-field variability surveys.\ {\it Monthly
Notices of the Royal Astronomical Society 356}, 557-567.
 
\refs Laughlin, G., Deming, D., Langton, J., Kasen, D., Vogt, S.,
Butler, P., Rivera, E., Meschiari, S.\ (2009) Rapid heating of the
atmosphere of an extrasolar planet.\ {\it Nature 457}, 562-564.
 
\refs L{\'e}ger, A., and 160 colleagues (2009) Transiting exoplanets
from the CoRoT space mission. VIII. CoRoT-7b: the first super-Earth
with measured radius.\ {\it Astronomy and Astrophysics 506}, 287-302.
 
\refs Lindberg, R., Stankov, A., Fridlund, M., Rando, N.\ (2009)
Current status of the assessment of the ESA Cosmic Vision mission
candidate PLATO.\ {\it Proceedings of the SPIE 7440}, 7440Z-1 to
7440Z-12

\refs Madhusudhan, N., Seager, S.\ (2009) A Temperature and Abundance
Retrieval Method for Exoplanet Atmospheres.\ {\it Astrophysical
Journal 707}, 24-39.

\refs Madhusudhan, N., Winn, J.~N.\ (2009) Empirical Constraints on
Trojan Companions and Orbital Eccentricities in 25 Transiting
Exoplanetary Systems.\ {\it Astrophysical Journal 693}, 784-793.

\refs Mandel, K., Agol, E.\ (2002) Analytic Light Curves for Planetary
Transit Searches.\ {\it Astrophys.\ J.\ 580}, L171-L175.
 
\refs Marzari, F., Nelson, A.~F.\ (2009) Interaction of a Giant Planet
in an Inclined Orbit with a Circumstellar Disk.\ {\it Astrophysical
Journal 705}, 1575-1583.
 
\refs Mazeh, T., and 19 colleagues (2000) The Spectroscopic Orbit of
the Planetary Companion Transiting HD 209458.\ {\it Astrophysical
Journal 532}, L55-L58.

\refs Mazeh, T., Zucker, S., Pont, F.\ (2005) An intriguing
correlation between the masses and periods of the transiting planets.\
{\em Mon.\ Not.\ Roy.\ Astr.\ Soc.\ 356}, 955-957.

\refs McCullough, P.~R., Stys, J.~E., Valenti, J.~A., Fleming, S.~W.,
Janes, K.~A., Heasley, J.~N.\ (2005) The XO Project: Searching for
Transiting Extrasolar Planet Candidates.\ {\it Publ.\ Astron.\ Soc.\
Pacific 117}, 783-795.
 
\refs Miller, N., Fortney, J.~J., Jackson, B.\ (2009) Inflating and
Deflating Hot Jupiters: Coupled Tidal and Thermal Evolution of Known
Transiting Planets.\ {\it Astrophysical Journal 702}, 1413-1427.
 
\refs Miralda-Escud{\'e}, J.\ (2002) Orbital Perturbations of
Transiting Planets: A Possible Method to Measure Stellar Quadrupoles
and to Detect Earth-Mass Planets.\ {\it Astrophys.\ J.\ 564},
1019-1023.

\refs Moutou, C., and 21 colleagues (2009) Photometric and
spectroscopic detection of the primary transit of the 111-day-period
planet HD 80606b.\ {\it Astronomy and Astrophysics 498}, L5-L8.
 
\refs Narita, N., Sato, B., Hirano, T., Tamura, M.\ (2009) First
Evidence of a Retrograde Orbit of a Transiting Exoplanet HAT-P-7b.\
{\it Publications of the Astronomical Society of Japan 61}, L35-L40.
 
\refs Nesvorn{\'y}, D.\ (2009) Transit Timing Variations for Eccentric
and Inclined Exoplanets.\ Astrophysical Journal 701, 1116-1122.

\refs Nutzman, P., Charbonneau, D.\ (2008) Design Considerations for a
Ground-Based Transit Search for Habitable Planets Orbiting M Dwarfs.\
{\it Publications of the Astronomical Society of the Pacific 120},
317-327.
 
\refs O'Donovan, F.~T., Charbonneau, D., Alonso, R., Brown, T.~M.,
Mandushev, G., Dunham, E.~W., Latham, D.~W., Stefanik, R.~P., Torres,
G., Everett, M.~E.\ (2007) Outcome of Six Candidate Transiting Planets
from a TrES Field in Andromeda.\ {\it Astrophys.\ J.\ 662}, 658-668.
 
\refs Olling, R.~P.\ (2007) LEAVITT: A MIDEX-class Mission for Finding
and Characterizing 10,000 Transiting Planets in the Solar
Neighborhood.\ ArXiv e-prints arXiv:0704.3072.

\refs P{\'a}l, A., Kocsis, B.\ (2008) Periastron precession
measurements in transiting extrasolar planetary systems at the level
of general relativity.\ {\it Monthly Notices of the Royal Astronomical
Society 389}, 191-198.

\refs Pepper, J., Gould, A., Depoy, D.~L.\ (2003) Using All-Sky
Surveys to Find Planetary Transits.\ {\it Acta Astronomica 53},
213-228.

\refs Pollacco, D.~L., and 27 colleagues (2006) The WASP Project and
the SuperWASP Cameras.\ {\it Publ.\ Astron.\ Soc.\ Pacific 118},
1407-1418.

\refs Pont, F., Zucker, S., Queloz, D.\ (2006) The effect of red noise
on planetary transit detection.\ {\it Monthly Notices of the Royal
Astronomical Society 373}, 231-242.

\refs Pont, F., Knutson, H., Gilliland, R.~L., Moutou, C.,
Charbonneau, D.\ (2008) Detection of atmospheric haze on an extrasolar
planet: the 0.55-1.05 $\mu$m transmission spectrum of HD 189733b with
the Hubble Space Telescope.\ {\em Mon.\ Not.\ Roy.\ Astr.\ Soc.\ 385},
109-118.

\refs Press, W.~H., Teukolsky, S.~A., Vetterling, W.~T., \& Flannery,
B.~P.\ (2007) {\it Numerical Recipes: The Art of Scientific
Computing}, 3rd ed., Cambridge University Press
 
\refs Queloz, D., and 39 colleagues (2009) The CoRoT-7 planetary
system: two orbiting super-Earths.\ {\it Astronomy and Astrophysics
506}, 303-319.

\refs Rabus, M., Alonso, R., Belmonte, J.~A., Deeg, H.~J., Gilliland,
R.~L., Almenara, J.~M., Brown, T.~M., Charbonneau, D., Mandushev, G.\
(2009) A cool starspot or a second transiting planet in the TrES-1
system? {\it Astronomy and Astrophysics 494}, 391-397.

\refs Ragozzine, D., Wolf, A.~S.\ (2009) Probing the Interiors of very
Hot Jupiters Using Transit Light Curves.\ {\it Astrophysical Journal
698}, 1778-1794.

\refs Reiger, S.~H.\ (1963) Starlight scintillation and atmospheric
turbulence.\ {\it Astronomical Journal 68}, 395.

\refs Rowe, J.~F., and 10 colleagues (2008) The Very Low Albedo of an
Extrasolar Planet: MOST Space-based Photometry of HD 209458.\ {\it
Astrophysical Journal 689}, 1345-1353.
 
\refs Russell, H.~N.\ (1948) The royal road of eclipses. {\it Harvard
Coll.\ Obs.\ Monograph 7}, 181-209.

\refs Ryan, P., Sandler, D.\ (1998) Scintillation Reduction Method for
Photometric Measurements.\ {\it Publications of the Astronomical
Society of the Pacific 110}, 1235-1248.
 
\refs Sackett, P.\ (1999) Searching for Unseen Planets via Occultation
and Microlensing, in {\em Planets Outside the Solar System: Theory and
Observations} (J.-M.\ Mariotti and D.\ Alloin, eds.), p.\ 189, Kluwer,
Dordrecht.

\refs Sasselov, D.~D.\ (2003) The New Transiting Planet OGLE-TR-56b:
Orbit and Atmosphere.\ {\it Astrophysical Journal 596}, 1327-1331.
 
\refs Sato, B., and 20 colleagues (2005) The N2K Consortium. II. A
Transiting Hot Saturn around HD 149026 with a Large Dense Core.\ {\it
Astrophys.\ J.\ 633}, 465-473.
 
\refs Seager, S., Hui, L.\ (2002) Constraining the Rotation Rate of
Transiting Extrasolar Planets by Oblateness Measurements.\ {\it
Astroph.\ J.\ 574}, 1004-1010.
 
\refs Seager, S., Mall{\'e}n-Ornelas, G.\ (2003) A Unique Solution of
Planet and Star Parameters from an Extrasolar Planet Transit Light
Curve.\ {\it Astroph.\ J.\ 585}, 1038-1055.

\refs Seager, S., Sasselov, D.~D.\ (2000) Theoretical Transmission
Spectra during Extrasolar Giant Planet Transits.\ {\it Astroph.\ J.\
537}, 916-921.

\refs Sing, D.~K., D{\'e}sert, J.-M., Lecavelier Des Etangs, A.,
Ballester, G.~E., Vidal-Madjar, A., Parmentier, V., Hebrard, G.,
Henry, G.~W.\ (2009) Transit spectrophotometry of the exoplanet
HD~189733b. I. Searching for water but finding haze with HST NICMOS.\
{\it Astronomy and Astrophysics 505}, 891-899.
 
\refs Snellen, I.~A.~G., de Mooij, E.~J.~W., Albrecht, S.\ (2009) The
changing phases of extrasolar planet CoRoT-1b.\ {\it Nature 459},
543-545.
 
\refs Southworth, J., Wheatley, P.~J., Sams, G.\ (2007) A method for
the direct determination of the surface gravities of transiting
extrasolar planets.\ {\em Mon.\ Not.\ Roy.\ Astr.\ Soc.\ 379},
L11-L15.
 
\refs Southworth, J.\ (2008) Homogeneous studies of transiting
extrasolar planets.\ I.\ Light-curve analyses.\ {\em Mon.\ Not.\ Roy.\
Astr.\ Soc.\ 386}, 1644-1666.

\refs Southworth, J.\ (2009) Homogeneous studies of transiting
extrasolar planets - II. Physical properties.\ {\it Monthly Notices of
the Royal Astronomical Society 394}, 272-294.

\refs Stello, D., and 24 colleagues (2009) Radius Determination of
Solar-type Stars Using Asteroseismology: What to Expect from the
Kepler Mission.\ {\it Astroph.\ J.\ 700}, 1589-1602.

\refs Sterne, T.~E.\ 1940.\ On the Determination of the Orbital
Elements of Eccentric Eclipsing Binaries.\ {\it Proc.\ Nat.\ Ac.\
Sci.\ 26}, 36-40.
 
\refs Struve, O.\ (1952) Proposal for a project of high-precision
stellar radial velocity work.\ {\it The Observatory 72}, 199-200.

\refs Swain, M.~R., Vasisht, G., Tinetti, G.\ (2008) The presence of
methane in the atmosphere of an extrasolar planet.\ {\it Nature 452},
329-331.
 
\refs Swain, M.~R., Vasisht, G., Tinetti, G., Bouwman, J., Chen, P.,
Yung, Y., Deming, D., Deroo, P.\ (2009) Molecular Signatures in the
Near-Infrared Dayside Spectrum of HD 189733b.\ {\it Astrophys.\ J.\
690}, L114-L117.

\refs Tamuz, O., Mazeh, T., Zucker, S.\ (2005) Correcting systematic
effects in a large set of photometric light curves.\ {\it Monthly
Notices of the Royal Astronomical Society 356}, 1466-1470.
 
\refs Tinetti, G., and 12 colleagues (2007) Water vapour in the
atmosphere of a transiting extrasolar planet.\ {\it Nature 448},
169-171.
 
\refs Torres, G., Winn, J.~N., Holman, M.~J.\ (2008) Improved
Parameters for Extrasolar Transiting Planets.\ {\it Astrophys.\ J.\
677}, 1324-1342.
 
\refs Torres, G., and 15 colleagues (2007) HAT-P-3b: A
Heavy-Element-rich Planet Transiting a K Dwarf Star.\ {\it Astrophys.\
J.\ 666}, L121-L124.
 
\refs Udalski, A., Paczynski, B., Zebrun, K., Szymanski, M., Kubiak,
M., Soszynski, I., Szewczyk, O., Wyrzykowski, L., Pietrzynski, G.\
(2002) The Optical Gravitational Lensing Experiment. Search for
Planetary and Low-Luminosity Object Transits in the Galactic
Disk. Results of 2001 Campaign.\ {\it Acta Astronomica 52}, 1-37.

\refs Vidal-Madjar, A., Lecavelier des Etangs, A., D{\'e}sert, J.-M.,
Ballester, G.~E., Ferlet, R., H{\'e}brard, G., Mayor, M.\ (2003) An
extended upper atmosphere around the extrasolar planet HD209458b.\
{\it Nature 422}, 143-146.
 
\refs Weldrake, D.~T.~F., Sackett, P.~D., Bridges, T.~J., Freeman,
K.~C.\ (2005) An Absence of Hot Jupiter Planets in 47 Tucanae: Results
of a Wide-Field Transit Search.\ {\it Astroph.\ J.\ 620}, 1043-1051.
 
\refs Winn, J.~N., Holman, M.~J., Roussanova, A.\ (2007a) The Transit
Light Curve Project.\ III.\ Tres Transits of TrES-1.\ {\it
Astrophysical Journal 657}, 1098-1106.

\refs Winn, J.~N., and 10 colleagues (2007b) The Transit Light Curve
Project.\ V.\ System Parameters and Stellar Rotation Period of HD
189733.\ {\it Astronomical Journal 133}, 1828-1835.

\refs Winn, J.~N., Henry, G.~W., Torres, G., Holman, M.~J.\ (2008)
Five New Transits of the Super-Neptune HD 149026b.\ {\it Astrophysical
Journal 675}, 1531-1537.

\refs Winn, J.~N., and 10 colleagues (2009a) On the Spin-Orbit
Misalignment of the XO-3 Exoplanetary System.\ {\it Astrophysical
Journal 700}, 302-308.

\refs Winn, J.~N., Johnson, J.~A., Albrecht, S., Howard, A.~W., Marcy,
G.~W., Crossfield, I.~J., Holman, M.~J.\ (2009b) HAT-P-7: A Retrograde
or Polar Orbit, and a Third Body.\ {\it Astrophysical Journal 703},
L99-L103.

\refs Young, A.~T.\ (1967) Photometric error
analysis. VI. Confirmation of Reiger's theory of scintillation.\ {\it
Astronomical Journal 72}, 747.

\refs Zahnle, K., Marley, M.~S., Freedman, R.~S., Lodders, K.,
Fortney, J.~J.\ (2009) Atmospheric Sulfur Photochemistry on Hot
Jupiters.\ {\it Astrophysical Journal 701}, L20-L24.